\documentclass[11pt,a4paper]{article}
\usepackage{jheppub}
\usepackage{amsmath}
\usepackage{multicol,bbm}
\usepackage{enumerate}
\usepackage{bibentry}
\usepackage{epstopdf}
\usepackage{upgreek}
\usepackage{amssymb}
\usepackage{bm}
\usepackage{multirow}
\usepackage{diagbox}
\usepackage{caption}
\usepackage{tabularx}

\def\be{\begin{equation}}
\def\ee{\end{equation}}
\def\ba{\begin{eqnarray}}
\def\ea{\end{eqnarray}}
\def\nl{\nonumber\\}

\def\a{\alpha}
\def\b{\beta}
\def\c{{\cdot}}
\def\d{\delta}
\def\ad{{\dot\alpha}}
\def\bd{{\dot\beta}}

\def\b#1{\overline{#1}}

\def\CP1{\mathbb{CP}^1}
\def\SL2C{\mathrm{SL}(2,\mathbb{C})}

\def\Z2{\mathbb{Z}_2}

\def\su2{{SU(2)}}

\def\a{{\alpha}}

\def\[{\left[}
\def\]{\right]}

\def\e{\epsilon}

\def\s{\sigma}
\def\a{\alpha}
\def\b{\beta}

\def\({\left(}
\def\){\right)}
\def\[{\left[}
\def\]{\right]}

\def\<{\langle}
\def\>{\rangle}

\def\i2{\frac{i}{2}}

\def\2F1{\,_2{\rm F}_1}

\begin{document}


\title{CHY formulae in 4d}


\author[a,b]{Yong Zhang}
\affiliation[a]{Department of Physics, Beijing Normal University, Beijing 100875, China}
\affiliation[b]{CAS Key Laboratory of Theoretical Physics, Institute of Theoretical Physics, Chinese Academy of Sciences, Beijing 100190, China}

\emailAdd{yongzhang@itp.ac.cn}

\date{\today}

\abstract{In this paper, we develop a rather general way to reduce integrands with polarization involved in the Cachazo-He-Yuan formulae, such as the reduced Pfaffian , its compactification and its squeezing, as well as the new object for $F^3$ amplitude. We prove that the reduced Pfaffian vanishes unless evaluated on a certain set of solutions. It leads us to build up the 4d CHY formulae using spinors, which strains off many useless solutions. The supersymmetrization is straightforward and may provide a hint to understand ambitwistor string in 4d.
}

\maketitle

\section{Introduction}

A new formulation for S-matrix of massless particles in arbitrary dimensions, dubbed as Cachazo-He-Yuan (CHY) formulation, has been developed for a large variety of theories~\cite{Cachazo:2013hca,Cachazo:2013iea,Cachazo:2014nsa,Cachazo:2014xea}. It expresses tree-level S-matrix as an integral over the moduli space of Riemann spheres, which are localized by a set of constraints, known as scattering equations~\cite{Cachazo:2013iaa, Cachazo:2013gna, Cachazo:2013hca}
\be
 {\cal E}_a:=\sum_{b\neq a} \frac{s_{a\,b}}{\sigma_{a}-\sigma_{b}} \,=\, 0,
  \quad \text{for}~a=1, 2, \ldots, n,
  \label{scatt}
\ee
where $s_{a\,b}=(k_a+k_b)^2=2 k_a\cdot k_b$, $\sigma_a$ denotes the position of the $a^{\rm th}$ puncture and we denote $\sigma_{a\,b}:=\sigma_a-\sigma_b$.It has been argued that what underpins the formulation is the ambitwistor string theory~\cite{Adamo:2013tsa,Mason:2013sva, Casali:2015vta}.

 The formulation has been inspired by Witten's revolutionary twistor string theory for ${\cal N}=4$ super-Yang-Mills theory (SYM) in four dimensions~\cite{Witten:2003nn}, and in particular the Roiban-Spradlin-Volovich-Witten (RSVW) formulae for all tree amplitudes in the theory~\cite{Roiban:2004yf}. Originally CHY discovered scattering equations in attempts to rewrite the equations in the delta functions of RSVW formulae without using 4d spinor helicity variables~\cite{Cachazo:2013iaa}, thus by construction they reduce to RSVW equations in four dimensions.
 More precisely, we have $n{-}3$ different sets of 4d equations, which are {\it polynomial} equations of degree $d=1,2,\ldots,n{-}3$. The $n{-}3$ sectors are labeled by $k'=d{+}1=2,\ldots, n{-}2$, which coincide with helicity sectors.
A set of equations, which are completely equivalent to RSVW equations, have been proposed in~\cite{Geyer:2014fka} based on ambitwistor string theory in four dimensions. It turns out that they are more convenient for our purposes, and in particular for helicity amplitudes. To write the equations in sector $k'$, we divide $n$ particles into two sets of $k'$ and $n{-}k'$ particles denoted as $-'$ and $+'$ respectively:
 \ba\label{4dr}
E_b^{\dot{\alpha}}\equiv\tilde\lambda_b^{\dot{\alpha}}  -t_b \sum_{p\in +'} \frac{t_p\tilde\lambda^{\dot{\alpha}}_p }{ \s_{b\,p}} \,=\, 0
  ~~{\text{for}}~b\in -',
  \qquad
 E_p^\alpha\equiv\lambda_p^\alpha - t_p\sum_{b\in -'} \frac{t_b\lambda_b^\alpha }{ \s_{p\,b}} \,=\, 0
  ~~{\text{for}}~p=+',\quad
\ea
 here the variables are $\sigma$'s and $t$'s, which can be combined into $n$ variables in $\mathbb{C}^2$, $\sigma^{\alpha}_a=\frac 1{t_a} (\sigma_a, 1)$.
 The $\sigma_{b\,p}$ is the abbreviation of $\sigma_b{-}\sigma_p$.
The $-'$ and $+'$ are arbitrary two sets of the $n$ external particles, with their length equal to $k'$ and $n-k'$ respectively. Different choices just correspond to different link representation \cite{Arkani-Hamed:2013jha,He:2012er}, which share the same solution of $\sigma$'s.
In this paper, we reserve $-$ and $+$ as the negative and positive helicity sets of external particles and $k$ the length of  $-$, {\it i.e.} the number of external particles of negative helicity. {\it A priori} there is no relation between solution sector and helicity sector.

We refer the readers to~\cite{He:2016vfi} for the direct derivation of \eqref{4dr} from \eqref{scatt}; in the same paper, it has been shown that \eqref{4dr} is equivalent to RSVW equations, and one can freely translate between  the two forms .  Each solution of \eqref{scatt} corresponds to a unique solution $\{\sigma_a, t_a\}$ of \eqref{4dr} for some $k'$, with identical cross-ratios of the $\sigma$'s. For each $k'$, \eqref{4dr} have an Eulerian number of solutions, $E_{n{-}3,k'{-}2}$, and the union of them for all sectors give $(n{-}3)!$ solutions of \eqref{scatt}, with $(n{-}3)!=\sum_{k'=2}^{n{-}2} E_{n{-}3,k'{-}2}$~\cite{Cachazo:2013iaa,Cachazo:2016sdc}.

%


It is highly non-trivial to reduce the localized integral measure of CHY formula, with delta functions of \eqref{scatt}, to that of 4d formula, with \eqref{4dr}, for some $k'$ sector. The reduction requires a sum over all sectors, and for each of them it results in a complicated conversion factor that depends on $k'$. In addition, after we  plug in spinor-helicity variables for {\it e.g.} Yang-Mills amplitudes, the CHY integrand behaves very differently in different helicity and solution sectors.
%
As we will see, the reduced Pfaffian plays the role of ``solution-filter": it is non-vanishing only on the solution sector that coincides with the helicity sector, which is why we have  a 4d formula for each helicity sector. What is even more interesting is that in the right sector, the polarization part of the CHY integrand exactly cancels the $k'$-dependent conversion factor from the measure! Thus two complications cancel out, and for YM we are left with a trivial Parke-Taylor factor in 4d.

Let's make the statement more precisely. For gauge theory and gravity, the most important ingredient is a $2n\times 2n$ skew matrix $\Psi_n$
{\small
\be\label{Pfa}
\Psi_n:=\left(
\begin{array}{cc}
A&-C^T\\
C&B
\end{array}
\right)\,;\quad 
A_{a\,b}=\begin{cases}
\frac{k_a\cdot k_b}{\sigma_{a\,b}}&a\neq b\\0&a=b
\end{cases},~
B_{a\,b}=\begin{cases}
\frac{\epsilon_a\cdot\epsilon_b}{\sigma_{a\,b}}&a\neq b\\0&a=b
\end{cases},~
C_{a\,b}=\begin{cases}
\frac{\epsilon_a\cdot k_b}{\sigma_{a\,b}}&a\neq b\\-\sum\limits_{c\neq a}C_{a\,c}&a=b
\end{cases},
\ee
}
and we define its reduced Pfaffian ${\rm Pf}{}'{\bm \Psi}_n:=\frac{(-)^{a{+}b}}{\sigma_{a\,b}} {\rm Pf} |\Psi_n|^{a\,b}_{a\,b}$ with $1{\leq} a{<}b{\leq} n$.

 We try to factorize the ${\rm Pf}{}'{\bm \Psi}_n$ into two parts depending on particles of negative and positive helicity respectively. Then we show in the right sector that is consistent to the helicity sector, each of the  parts combines to a reduced determinant while in other sector one of the part must vanish. That is,
   \ba\label{id1}
{\mathrm{Pf'}{\bm\Psi}_n\big|}_{k'}=\d_{kk'} \mathrm{det'}\,
{\bm h}_k\;\mathrm{det'}\,{\bm {\tilde{h}}}_{n-k}\;,
\ea
Here the two matrices, the  $k\times k$ matrix $\bm{h}_k$ and $(n{-}k)\times (n{-}k)$ one $\bm{\tilde{h}}_{n-k}$  essentially introduced in~\cite{Geyer:2014fka} (see also \cite{Cachazo:2012kg,Cachazo:2012pz}) are given by
\ba\label{haa}
h_{ab}=\frac{\<ab\>}{\s_{ab}} \quad{\text {for}}~ a\neq b,
\quad h_{aa}=-\sum_{b\neq a}\frac{t_b}{t_a}h_{ab}\quad a,b\in -\;,\nl
\tilde{h}_{ab}=\frac{[ab]}{\s_{ab}} \quad{\text {for}}~ a\neq b,
\quad \tilde{h}_{aa}=-\sum_{b\neq a}\frac{t_b}{t_a}\tilde{h}_{ab}\quad
a,b\in+\;,
\ea
and we define ${\det}'\bm{h}_k=\det |\bm{h}_k|^a_b/(t_a t_b)$ (similarly for ${\det}'  \bm{\tilde{h}}_{n-k}$) where we use $|\bm{h}_k|^a_b$ to denote the minor with any row $a$ and column $b$ deleted.

We rearrange the $\mathrm{Pf'}{\bm\Psi}_n$ using some fundamental gauge invariant or almost gauge invariant objects. It is either a (modified) trace of linearised field strength  ornamented with some $\s$'s or $C_{aa}$.
We view the 4d scattering equations \eqref{4dr} as a change of variables: we refer to $\lambda_{b\in -'}$, $\tilde\lambda_{p\in +'}$ and $t_a, \sigma_a$ as ``data" and the 4d scattering equations \eqref{4dr} as writing  $\tilde\lambda_{b\in -'}$ and $\lambda_{p\in+'}$ in terms of the data. After plugging in this change of variables, the $C_{aa}$ in ${\bm \Psi}_n$ directly reduces to object made up of spinors.  What left to do is to deal with all kind of trace. After all, somehow, we find the reduced Pfaffian reduces to the two  reduced determinants. This way of reduction is rather general: not only the reduced Pfaffian, but also many other integrands, such as the reduced compactified Pfaffian used in Einstein-Maxwell, Yang-Mills-Scalar, Dirac-Born-Infeld amplitudes or the new object ${\cal P}_n$  used in $F^3,R^2,R^3$ amplitudes are also related  to these two (extended) matrices. It may even be applied at loop level \cite{He:2016mzd}.

The paper is organized as follows. In section 2, we introduce the CHY formulae in 4d. In section 3, we study the reduction of Pfaffians to 4d for $k'{=}k$. First we see how ${\rm Pf}{\bm {\Psi}}_n$ factorizes in 4d in a manifeslty gauge-invairant way, which naturally leads to the 4d matrices ${\bm h}_k$ and  ${\bm {\tilde h}}_{n-k}$. Then we present the beautiful reduction of ${\rm Pf'}{\bm {\Psi}}_n$ , in a similar but more non-trivial way. In section 4,  we move to general case with arbitrary $k'$, which requires generalized version of ${\bm h}_k^{k'}$ and  ${\bm {\tilde h}}_{n-k}^{k'}$ matrices. We show that both ${\rm Pf'}{\bm {\Psi}}_n$ and ${\cal P}_n$ reduce nicely into the generalized ${\bm h}_k^{k'}$ and  ${\bm {\tilde h}}_{n-k}^{k'}$; while  ${\rm Pf'}{\bm {\Psi}}_n$ directly vanishes when $k'\neq k$, ${\cal P}_n$ does not and gives interesting formulae in 4d. The reduction of the reduced compactified Pfaffian and squeezed Pfaffian is put in Appendix \ref{compactified},\ref{squeezed}.

\section{4d CHY formulae}
 We start with  CHY formula for tree-level S-matrix of $n$ massless particles:
\be\label{int1}
M_n=\frac 1 {{\rm vol~SL}(2,\mathbb{C})}\,\int\,\prod_{a=1}^n d\,\sigma_a~\prod_{a=1}^n {}' \,\delta({\cal E}_a)~{\cal I}_n(\{\sigma, k, \ldots\})=\sum_{\rm solutions} \frac{{\cal I}_n(\{\sigma, k, \ldots\})}{\det' \Phi_n }\,,
\ee
where the precise definition of the integral measure including delta functions can be found in~\cite{Cachazo:2013hca}, and ${\cal I}_n$ is the CHY integrand depending on the theory.  In the second equality one sums over $(n{-}3)!$ solutions of \eqref{scatt}, evaluated on the integrand and the Jacobian, which is defined as a reduced determinant:
\be\label{Phi}
\det{}' \Phi_n:=\frac{\det |\Phi_n|^{p\,q\,r}_{a\,b\,c}}{| p\,q\,r| |a\,b\,c|}\,\quad {\rm with} \qquad \Phi_{a\,b}=\frac{s_{a\,b}}{\sigma_{a\,b}^2}\,,\quad {\rm for}~a\neq b\,,\qquad \Phi_{a\,a}=-\sum_{b\neq a} \Phi_{a\,b}\,,
\ee
where the $n\times n$ matrix $(\Phi_n)$, with entries $\{\Phi_{a\,b}\}:=-\partial\{{\cal E}_a\}/\partial\{\sigma_b\}$, is the derivative matrix; the rows $p,q,r$ and columns $a, b, c$ are deleted (corresponding to deleted equations and variables, respectively), and we have two Fadeev-Popov factors, defined as $| a\,b\,c|:=\sigma_{a\,b}\sigma_{b\,c}\sigma_{c\,a}$ .

For gauge theory and gravity, the most important ingredient is the reduced Pfaffian ${\rm Pf}' {\bm \Psi}_n$ given in \eqref{Pfa}. Many other integrands can be abtained by doing some operation on it.
 The CHY integrand for $n$-point Yang-Mills tree amplitudes reads
\be
{\cal I}_n^{\rm YM}={\cal C}_n~{\rm Pf}'{\bm \Psi}_n\,,\qquad {\cal C}_n=\frac{{\rm Tr} (T^{I_1}~T^{I_2}\cdots T^{I_n})}{\sigma_{1\,2}~\sigma_{2\,3}\cdots \sigma_{n\,1}}~+~{\rm permutations}\,,
\ee
where ${\cal C}_n$ is the color-dressed Parke-Taylor factor, with the sum over $(n{-}1)!$ inequivalent permutations.

The general 4d formula in solution sector $k'$ for $n$-point amplitudes reads:
\be\label{int2}
M_{n,k'}=\frac 1 {{\rm vol~GL}(2,\mathbb{C})}\,\int\,\prod_{a=1}^n d^2 \sigma_a\,\prod_{\substack {b\in -'\\p\in +'}}{}'\,\delta^2 (E_b)~\delta^2(E_p)~I_{n}(\{\sigma, \lambda, \tilde\lambda\})=\sum_{k'-{\rm sec.~sol.}} \frac{I_{n} (\{\sigma, \lambda, \tilde\lambda\})}{J_{n,k'}}
\ee
where $d^2\sigma_a:=d \sigma_a\,\frac {d t_a}{t_a}$, and in addition to 4 deleted variables by GL(2), 4 redundant equations in \eqref{4dr} are deleted which give overall delta functions for momentum conservation. In the second equality, one first sums over the Eulerian number, $E_{n{-}3,k'{-}2}$, solutions in sector $k'$. The $J_{n,k'}$ is the Jacobian of the localized $2n-4$ integrals
\be\label{J}
J_{n,k'}=\big(\prod_{a=1}^nt_a\big)\frac{{\rm det} \left(\partial \{ E^{\dot\alpha}_{b\neq c,d}, E^{\alpha}_p \}/\partial \{t_{a\neq m}, \sigma_{a \neq u,v,w}\}\right)} {t_m\,\sigma_{u,v}\,\sigma_{v,w}\,\sigma_{w,u}~\langle c\,d\rangle^2}
\ee
where we have chosen to eliminate $t_m, \sigma_u, \sigma_v, \sigma_w$ and $E^{\dot\alpha}_{b=c,d}$, with the FP factor $\langle c\,d\rangle^2$ (for $E^{\alpha}_{p\neq q,r}$ the FP factor is $[q\,r]^2$).

The relation between the two Jacobians is simple.  Viewing \eqref{4dr} as a change of variables and  plugging in it , we find
\ba\label{iden2}
\det{}' \Phi_n (\{s_{a b}, \sigma_a\}) |_{k'}&=  J_{n,k'} ~\mathrm{det'}\,{\bm{h}}_{k'}\mathrm{det'}\,{\bm{\tilde{h}}}_{n-k'}\,.\label{id2}
\ea
Here we don't need to plug in any solutions, but simply make a change of variables, so this is really an equality between rational functions of the data, {\it i.e.} $\lambda_b$'s, $\tilde\lambda_p$'s, and $\sigma_a, t_a$'s. The two reduced determinants  $\mathrm{det'}\,{\bm{h}}_{k'}$ and $\mathrm{det'}\,{\bm{\tilde{h}}}_{n-k'}$  can be thought as  two resultants and are divided by $\det{}' \Phi_n$ as discussed in~\cite{Cachazo:2013zc}. We find that the quotient is just $J_{n,k'} $. A conjecture about the closed form of  $J_{n,k'} $ is put in Appendix \ref{jnk}.

Thanks to \eqref{id1},  for gluon amplitudes, the integrand is nothing but the (color-dressed) Parke-Taylor factor $I^{\rm YM}_n={\cal C}_n$. Different from \eqref{int1}, any $\frac{\s_{bc}}{t_bt_c}\frac{\s_{pq}}{t_pt_q}$ or $\frac{\s_{bp}}{t_bt_p}$ with $b,c\in -'$ and $p,q\in +'$ is ${\rm GL}(2,\mathbb{C})$ invariant and any known 4d integrand added with these objects could be a new 4d integrand, for example we add some $\frac{\s_{bp}}{t_bt_p}$ to the $I^{\rm YM}_n$ and we get those for QCD in \cite{He:2016iqi}.

In this paper, we  explicitly demonstrate the first identity \eqref{id1}. Compared to this identity, the second one \eqref{id2} is a more boring one, as there is no polarization involved and just kinematics reducing to $4$ dimensions. One can check as many points as we want, without any difficulties (we have checked up to $50$ points with all solution sectors numerically). A proof based on direct inspection should be straightforward.

\section{Reduced Pfaffian in 4d for the $k'=k$ sector}\label{Pfaffian}

In this section, we show in a constructive way how the reduced Pfaffian factorizes in four dimensions for the solution sector that coincides with its helicity sector, $k'=k$. We will proceed in two steps: as a warm up, we show how it works for the vanishing Pfaffian ${\rm Pf} \bm{\Psi}_n$, which factorizes into two vanishing determinants in 4d; then we apply it to the more non-trivial case of the reduced Pfaffian and show ${\rm Pf}' \bm{\Psi}_n=\det' {\bm h}~\det' {\bm {\tilde h}}$. The reason for doing so is that both ${\rm Pf} \bm{\Psi}_n$ and ${\rm Pf}' \bm{\Psi}_n$ have similar expansions, as first studied in~\cite{Lam:2016tlk}, and we review them here.

 From the definition of Pfaffian and thanks to the special structure of $2n \times 2n$ matrix $\bm{\Psi}_n$, we can expand ${\rm Pf}\bm{\Psi}_n$ as a sum over $n!$ permutations of labels $1,2,\ldots, n$, denoted as $p\in S_n$
\ba\label{psi1}
{\rm Pf}\,\bm{\Psi}_n=\sum_{p\in S_n}\,{\rm sgn}(p)\,\Psi_p=\sum_{p\in S_n}\,{\rm sgn}(p)\,\Psi_I\Psi_J\cdots\Psi_K,
\ea
where ${\rm sgn}(p)$ denotes the signature of the permutation $p$ and in the second equality, we use the unique decomposition of any permutation $p$
into disjoint cycles $I,J,\cdots,K$ given by
\ba
I=(a_1a_2\cdots a_i),\quad J=(b_1b_2\cdots b_j),\cdots ,\quad K=(c_1c_2\cdots c_k)\;;
\ea
each $\Psi_p$ is the product of its ``cycle factors" $\Psi_I\Psi_J\cdots\Psi_K$, which we define now. When the length of a cycle equals one, its cycle factor $\Psi_{(a)}$ is given by the diagonal of $C$-matrix:
\ba\label{psii}
\Psi_{(a)}:=C_{aa}=-\sum_{b\neq a}\frac{\e_a\c k_b}{\s_{ab}}\;,
\ea
and when the length exceeds one {\it e.g.} $i>1$, the cycle factor is given by
\ba\label{psi2}
\Psi_I=\Psi_{(a_1a_2\cdots a_i)}:=\frac{\frac{1}{2}~\mathrm{tr}(f_{a_1}f_{a_2}\cdots f_{a_i})}{\s_{(a_1a_2a_3\cdots a_i)}}
\qquad \mathrm{with}\quad f_a^{\mu\nu}=k_a^{\mu}\e_a^{\nu}-\e_a^{\mu} k_a^{\nu}\;.
\ea
Here $\s_{(a_1a_2a_3\cdots a_i)}=\s_{a_1a_2}\s_{a_2a_3}\cdots\s_{a_ia_1}$. The trace is over Lorentz indices and $f^{\mu\nu}$ is the linearized field strengths of gluons. Note that the decomposition is manifestly gauge invariant: for cycle factors with length more than 1 \eqref{psi2}, the trace of $f^{\mu\nu}$ is gauge invariant, while for $1$-cycles, \eqref{psii}, the factor is gauge invariant on the support of scattering equations ~\eqref{scatt}.

The reduced Pfaffian ${\rm Pf'}{\bm \Psi}_n$, as discussed in~\cite{Lam:2016tlk}, is  different from ${\rm Pf}{\bm \Psi}_n$. Because the $1^{\rm th},n^{\rm th}$ columns and rows have been deleted, the numerator of the cycle containing $1$ and $n$ becomes $\frac{1}{2} \e_{1}\c \big( f_{a_2}f_{a_3}\cdots f_{a_{i-1}}\big)\c \e_{n}$ instead of a trace. Then
\ba\label{pin}
{\rm Pf}'{\bm \Psi}_n=\sum_{p\in S_n}\!^\prime {\rm sgn}(p)\,W_I\Psi_J\cdots\Psi_K\;,
\ea
with
\ba\label{opencycle}
W_I=W_{[1 a_2\cdots a_{i{-}\!1} n]}=\frac{\frac{1}{2} \e_{1}\c \big( f_{a_2}f_{a_3}\cdots f_{a_{i-1}}\big)\c \e_{n}}{\s_{(1a_2a_3\cdots a_{i-1}n)}}\;.
\ea
Here $I,J,\cdots K$ are the cycles of permutation $p$. The prime on the summation sign indicates that the sum is taken over all $p\in S_n$ such that $1$ is changed into $n$. There are $(n-1)!$ such permutations in $S_n$ so the sum consists of $(n-1)!$ terms.

The key observation in~\cite{Lam:2016tlk} allows us to expand the reduced Pfaffian in terms of building blocks, each of which is either the product of various closed cycles or an open cycle involving the two deleted labels.
Closed cycles have a very good property that they won't contribute unless all of their elements belong to same helicity.
While the open cycle is much tougher, as it's not gauge invariant individually (dependent on the gauge of the two deleted particles) and wont't vanish when their elements come from different helicity sets.  As a warm up, we show in the first subsection the Pfaffian ,which is the product of only closed cycles~\cite{Lam:2016tlk} ,factorizes . Though  the Pfaffian equals zero, it very non-trivially factorizes  into determinants of two matrices. Also it is the naturel way to introduce the two matrices $\bm{h}_k$ and  $\bm{\tilde{h}}_{n-k}$ \eqref{haa}. In the next subsection we carefully deal with the open cycle and finally factorize the reduced Pfaffian to two reduced determinants.


\subsection{The Pfaffian in 4d}
Let's start with the Pfaffian, ${\rm Pf}\bm{\Psi}_n$. In 4 dimension, $f^{\mu\nu}$ reduces to a self-dual part and an anti-self-dual  part: $f^{\mu\nu}\rightarrow \e^{\a\b}f^{\ad\bd}+\e^{\ad\bd}f^{\a\b}$. We denote these two parts as $f^-$ and $f^+$ respectively. An important property is that
any two adjoint linearised strength fields $f_b^- \,f_p^+$ in the trace can exchange their place if the helicity of $b,p$ are different,
i.e.
\ba
\cdots f_b^- f_p^+\cdots=\cdots f_p^+ f_b^-\cdots\;.
\ea
So we can always reduce those traces where particles of negative or positive helicity are mixed each other to split ones which have a simple reduction in 4d. Then
\be
\mathrm{tr}~(f_{a_1}\,f_{a_2} \cdots  f_{a_i})=\begin{cases}
2~\langle a_1a_2\rangle~\langle a_2 a_3\rangle \cdots \langle a_i a_1\rangle\,, \qquad &\{a_1,a_2,\cdots a_i\}\subset -\\
2~[a_i a_{i-1}]~[a_{i-1}a_{i-2}] \cdots [a_1 a_i]\,, \qquad &\{a_1,a_2,\cdots a_i\}\subset +\\
\langle b_1b_2\rangle \cdots \langle b_x b_1\rangle
 [p_{y} p_{y-1}]\cdots [p_1 p_y]
\,,& \quad {\rm otherwise}
\end{cases}
\;,
\ee
Here $b_1,b_2,\cdots,b_x$ are all the particles of negative helicity from $a_1,a_2,\cdots,a_i$ with their ordering unchanged and similarly $p_1,p_2,\cdots,p_y$ are all the particles of positive helicity from $a_1,a_2,\cdots,a_i$ with their ordering unchanged. Note that $\mathrm{tr}~(f_{a_1}\,f_{a_2} \cdots  f_{a_i})$  directly vanishes if there is only  one particle of negative helicity or  only one particle of positive helicity in $a_1,a_2,\cdots,a_i$. However we see that the remaining case still effectively vanish as we always add up all permutations (see \eqref{psi1}) while
\ba
\sum_{\{\a\}\in {\rm OP}(\{b_1,b_2,\cdots,b_x\},\{p_1,p_2,\cdots,p_y\})}\frac{1}{\s_{(\{\a\})}}=0\;.
\ea
Here the sum is over ordered permutations $``$OP$"$, namely permutations of the labels in the joined set $\{b_1,b_2,\cdots,b_x\},\{p_1,p_2,\cdots,p_y\}$
such that the ordering within $\{b_1,b_2,\cdots,b_x\}$ and $\{p_1,p_2,\cdots,p_y\}$ is preserved. Therefore, in the sum of \eqref{psi1}, we can effectively write $\mathrm{tr}~(f_{a_1}\,f_{a_2} \cdots f_{a_i})$ in 4d in a remarkably simple way:
\be\label{tr}
\frac 1 2 \mathrm{tr}~(f_{a_1}\,f_{a_2} \cdots f_{a_i}) \to \begin{cases}
\langle a_1a_2\rangle~\langle a_2 a_3\rangle \cdots \langle a_i a_1\rangle\,, \qquad &\{a_1,a_2,\cdots a_i\}\subset -\\
[a_1 a_2]~[a_2a_3] \cdots [a_i a_1]\,, \qquad &\{a_1,a_2,\cdots a_i\}\subset +\\
0\,, \qquad &{\rm otherwise}
\end{cases}
\;,
\ee

Motivated by \eqref{tr}, we recall the off-diagonal elements of the $k\times k$ matrix $\bm{h}_k$ and $(n{-}k)\times (n{-}k)$ one $\bm{\tilde{h}}_{n-k}$ essentially introduced in~\cite{Geyer:2014fka} (see also~\cite{Cachazo:2012kg,Cachazo:2012pz}):
\ba\label{offdiagonal}
h_{ab}=\frac{\<ab\>}{\s_{ab}} \quad a\neq b,\;a,b\in -\;,
\qquad
\tilde{h}_{ab}=\frac{[ab]}{\s_{ab}}\ \ \quad a\neq b,\;a,b\in +\;.
\ea
It is clear that when we have any cycle factor with length at least 2, it must be given by the chain product of such off-diagonal elements
\ba\label{psiI}
\Psi_{(a_1a_2\cdots a_i)}\rightarrow
\begin{cases}
h_{a_1a_2}h_{a_2a_3}\cdots h_{a_i a_1} \qquad &\{a_1,a_2,\cdots a_i\}\subset -\\
\tilde{h}_{a_1a_2}\tilde{h}_{a_2a_3}\cdots \tilde{h}_{a_i a_1} \qquad &\{a_1,a_2,\cdots a_i\}\subset +\\
0 &{\rm otherwise}
\end{cases}
\;,
\ea
To this point we have not used scattering equations and solution sectors in four dimensions. The  non-trivial part of the reduction concerns 1-cycle, or the diagonal entries of $C$-matrix. Note that $\Psi_{(a)}=C_{aa}$ is only gauge invariant on the support of scattering equations, so it is not surprising that to reduce it nicely one needs to use scattering equations in four dimensions.
 Now we derive the explicit expression of $C_{aa}$ .
When $a\in -$ and  $a\in -'$, we have
\ba
C_{aa}^-=-\sum_{{b\in {\rm -'},\,b\neq a}}\frac{\<ab\>[b\mu]}{[a\mu]\s_{ab}}
-\sum_{p\in {\rm +'}}\frac{\<ap\>[p\mu]}{[a\mu]\s_{ap}}\;.
\ea
Note that $C_{aa}$ depends on $\s$ and because of the 4d scattering equations \eqref{4dr}, we can make  the change of variables
 \begin{align}\label{changes}
\tilde\lambda_b^{\dot{\alpha}}  =t_b \sum_{p\in +'} {t_p\tilde\lambda^{\dot{\alpha}}_p \over \s_{b\,p}} \
  ~~\text{for}~b\in -',
  \qquad
\lambda_p^\alpha =t_p\sum_{b\in -'} {t_b\lambda_b^\alpha \over \s_{p\,b}} \
  ~~\text{for}~p=+',
\end{align}
Such that $C_{aa}^-$ reduces to:
\ba\label{a21}
C_{aa}^-&=&-\frac{1}{[a\mu]}
\sum_{b\neq a;\,p}
\Big(\frac{\<a b\>t_bt_p[p\mu]}{\s_{bp}\s_{ab}}+\frac{\<a b\>t_bt_p[p\mu]}{\s_{pb}\s_{ap}}
\Big)\nl
&=&-\frac{1}{[a\mu]}
\sum_{b\neq a;\,p}
\frac{\<a b\>t_bt_p[p\mu]}{\s_{bp}}\big(\frac{1}{\s_{ab}}-\frac{1}{\s_{ap}}\big)\nl
&=&-\frac{1}{[a\mu]}
\sum_{b\neq a;\,p}
\frac{\<a b\>t_bt_p[p\mu]}{\s_{ab}\s_{ap}}\;.
\ea
In the last equality, we have collected the denominators together such that $\s_{bp}$ is canceled.  Now $C_{aa}^-$ factorizes into two factors
\ba\label{a22}
C_{aa}^-=-\left(\sum_{b\neq a}
\frac{t_b\<a b\>}{t_a\s_{ab}}\right)\left(\sum_{p}\frac{t_at_p[p\mu]}{\s_{ap}[a\mu]}\right)=-\sum_{{b\in {\rm -'},\,b\neq a}}
\frac{t_b\<a b\>}{t_a\s_{ab}}\;.
\ea
 All gauge dependence is in the latter factor
and it can be eliminated by scattering equations as $ t_a \sum_{p\in +'} {t_p\tilde\lambda^{\dot{\alpha}}_p \over \s_{a\,p}}=\tilde\lambda_a^{\dot{\alpha}} $ ~\eqref{changes}.

Similarly we can work out the case of $a\in +$ and $a\in +'$
\ba\label{cii2}
C_{aa}^+=-\sum\limits_{b\neq a;\ b\in{\rm +'}}\frac{t_b}{t_a}\frac{[ab]}{\s_{ab}}\;.
\ea
We first discuss the $k'=k$ case and without loss of generality let's consider $-'{=} -$, which makes our discussion simpler. Then the above two cases are already enough here , postponing other two cases in the following sections. Miraculously, $C_{aa}$ reduces to diagonal entries of $\bm{h}_k$ or $\bm{\tilde{h}}_{n-k}$~\cite{Geyer:2014fka} depending on the helicity:
\ba
h_{aa}=C_{aa}^-=-\sum_{\tiny{\substack{b\neq a\\b\in -}}}\frac{t_b}{t_a}\frac{\<ab\>}{\s_{ab}}\quad a\in-\;,
\qquad
\tilde{h}_{aa}=C_{aa}^+=-\sum_{\tiny{\substack{b\neq a\\b\in +}}}\frac{t_b}{t_a}\frac{[ab]}{\s_{ab}}\quad a\in+\;.
\ea
The important thing is that the diagonal entry is a linear combination of off-diagonal entries in that row/column. With these diagonal entries of $\bm{h}_k$ or $\bm{\tilde{h}}_{n-k}$, the reduction for $\Psi_{(a_1a_2\cdots a_i)}$ with $i>1$ or $i=1$ (for $k'=k$) are both spelled out in one nice formula, \eqref{psiI}.

We find $h_{a_1a_2}h_{a_2a_3}\cdots h_{a_ia_1}$ in \eqref{psiI} is just the ingredient of
 $\mathrm{det}\,{\bm h}_{k}$,
\ba\label{horigin}
\mathrm{det}\;\bm{h}_k=\sum_{q\in S_k}{\rm sgn}(q) h_{I_1}h_{I_2}\cdots h_{I_s}\;,\quad {\rm with}\  h_{I}=h_{(a_1a_2\cdots a_i)}=h_{a_1a_2}h_{a_2a_3}\cdots h_{a_ia_1}\;,
\ea
where the sum is over all permutations of particles of negative helicity, {\it i.e.} $q\in S_k$ and  $I_1,I_2,\cdots,I_s$ are the cycles of the permutation $q$. Similarly works for $\tilde{h}_{a_1a_2}\tilde{h}_{a_2a_3}\cdots \tilde{h}_{a_i a_1}$.

Then, we see that ${\rm Pf}{\bm \Psi}_n$ factorizes to two parts depending on particles of negative or positive helicity respectively, with most of the terms vanishing and the surviving terms combining to $\mathrm{det}\,{\bm h}_{k}$ or $\mathrm{det}\,{\bm {\tilde{h}}}_{n-k}$,
\ba\label{id0}
{{\rm Pf}{\bm \Psi}_n\big|}_{k'=k}=\mathrm{det}\,{\bm h}_{k}\,\mathrm{det}\,{\bm {\tilde{h}}}_{n-k}\;.
\ea
Obviously both $\mathrm{det}\,{\bm h}_{k}$ and $\mathrm{det}\,{\bm {\tilde{h}}}_{n-k}$ vanish since they both have a null vector; this is consistent with the fact that ${\rm Pf}{\bm \Psi}_n$ vanishes due to the two null vectors.
\subsection{reduced Pfaffian in $4$ dimensions}

Now we turn to ${\rm Pf}'{\bm \Psi}_n$. Now we need to deal with the open cycle. Similarly,
 we can always reduce these mixed open brackets into split one as  any two adjoint linearised strength fields $f_b^- \,f_p^+$ in the kinematic numerator of open brackets $\e_1\cdots f_b^- f_p^+\cdots\e_n$ can exchange their place if the helicity of $b,p$ are different,
i.e.
\ba
\e_1\cdots f_b^- f_p^+\cdots\e_n=\e_1\cdots f_p^+ f_b^-\cdots\e_n\;.
\ea
Note that this equality is true no matter what the helicity of $1$ and $n$ are. In the following demonstration we need to delete two columns and rows from negative and positive helicity set respectively, so we assign $1^-$  and $n^+$.  Using this property, we can always rearrange the kinematic numerator in a split form with the ordering of particles of negative helicity and the ordering of particles of positive helicity unchanged respectively. For example, with $n{>}6$,
\ba
\e_1^-\c f_{5}^+f_{2}^-f_{3}^-f_{6}^+f_{4}^-\c \e_n^+=\e_1^-\c f_{2}^-f_{5}^+f_{3}^-f_{6}^+f_{4}^-\c \e_n^+
=\cdots&=&\e_1^-\c f_{2}^-f_{3}^-f_{4}^-f_{5}^+f_{6}^+\c \e_n^+\nl
&=&
\frac{2\, \langle 12\rangle  \langle 23\rangle\langle 34\rangle\langle4\mu \rangle  [\mu5
   ] [56]
   [6n]}{ [1\mu ]\langle n\mu \rangle }.\qquad
\ea
 All ${3 \choose 5}=10$ such kinematic numerators of open cycles whose ordering of negative and positive particles between $1$ and $n$ are $2,3,4$ and $5,6$ respectively equal to  $\e_1^-\c f_{2}^-f_{3}^-f_{4}^-f_{5}^+f_{6}^+\c \e_n^+$. Further on, all such kinematic numerator can reduce to a product of  some simple angle brackets and square brackets as shown in the last equality. Here $|\mu],|\mu\>$ are the reference of 1,n respectively.

For the general case with $x$ particles of negative and $y$ particles of positive helicity  between $1$ and $n$,
 there are ${x \choose x+y}$ cycles whose kinematic numerators are equal to those of a certain split open cycles and they all reduce to a product of  some simple angle brackets and square brackets,
\ba\label{open1}
\frac{1}{2} \e_1^-\c f_{b_1}^-f_{b_2}^-\c\c f_{b_x}^-f_{p_y}^+f_{p_{y-\!1}}^+\c\c f_{p_1}^+\c\e_n^+=
\frac{\<1b_1\>\<b_1b_2\>\c\c\<b_{x-\!1}b_x\>\<b_{x}\mu\>\,[\mu p_y][p_y p_{y-\!1}]\c\c [p_2p_1][p_1n]}{[1\mu]\langle n\mu \rangle }\;.\qquad
\ea
 Here $|\mu],|\mu\>$ are the reference of 1,n respectively, {\it i.e.} $\e_1^-=\frac{|1\>[\mu|}{[1\mu]},\e_n^+=\frac{|n]\<\mu|}{\<n\mu\>}$ and we have used the reversed ordering $p_y,p_{y-\!1},\cdots,p_1$ for later convenience.

Since ${x \choose x+y}$ such open brackets share same kinematic numerator, we try to combine their denominators. They happen to be combined to the partial fraction identity (analogous to Kleiss-Kuijf relations of amplitudes),
\ba
(-)^{|\rho|}\sum_{\{\a\}\in {\rm OP}(\{\b\},\{\rho^T\})}\frac{1}{\s_{(1,\{\a\},n)}}=\frac{1}{\s_{(1,\{\b\},n,\{\rho\})}}\;,
\ea
 here $\{\a\}$ means $a_2,a_3,\cdots,a_{i-1}$ and $\{\b\}$, $\{\rho\}$ means  $b_1,b_2,\cdots,b_x$, and $p_1,p_2,\cdots,p_y$ respectively.  $\{\rho^T\}$ denotes the reverse ordering of the labels $\{\rho\}$ .

Then  $(-)^{|\rho|}\sum_{\{\a\}\in {\rm OP}(\{\b\},\{\rho^T\})}\Psi_{[1 a_2\cdots a_{i{-}\!1} n]}$ combines to
\begin{align}\label{delete2k}
\frac{\frac{1}{2} \e_1^-\c f_{b_1}^-f_{b_2}^-\c\c f_{b_x}^-f_{p_y}^+f_{p_{y-\!1}}^+\c\c f_{p_1}^+\c\e_n^+}{ \s_{(1b_1\c\c b_x n p_1 p_2\c\c p_y)}}
=&
\left(h_{1b_1}h_{b_1b_2}\c\c h_{b_{x-1}b_x}\frac{\<b_x\mu\>}{\<n\mu\>\s_{b_x n}}\right)\,\left(\tilde{h}_{np_{1}}\tilde{h}_{p_1p_{2}}\c\c \tilde{h}_{p_{y-1}p_y }\frac{[\mu p_y]}{[1\mu]\s_{1p_y}}\right)\nl
:=&h_{[1b_1b_2\cdots b_x]}\tilde{h}_{[np_1p_2\cdots p_y]}
\end{align}
In the first equality, we have plugged in \eqref{open1}. In the second equality, we have defined $h_{[1b_1b_2\cdots b_x]}$  as $h_{1b_1}h_{b_1b_2}\c\c h_{b_{x-1}b_x}\frac{\<b_x\mu\>}{\<n\mu\>\s_{b_x n}}$  and $\tilde{h}_{[np_1p_2\cdots p_y]}$ as
$\tilde{h}_{np_{1}}\tilde{h}_{p_1p_{2}}\c\c \tilde{h}_{p_{y-1}p_y }\frac{[\mu p_y]}{[1\mu]\s_{1p_y}}$.
 Here we can treat $1$ as $b_0$ and if there is no particles of negative helicity between $1$ and $n$, $\frac{\<b_x\mu\>}{\<n\mu\>\s_{b_x n}}$ reduces to $\frac{\<1\mu\>}{\<n\mu\>\s_{1n}}$; Similarly we can treat $n$ as $p_0$ and if there is no particles of positive helicity between $1$ and $n$, $\frac{[\mu p_y]}{[1\mu]\s_{1p_y}}$ reduces to $\frac{[\mu n]}{[1\mu]\s_{1n}}$.
Note that these prefactors  $\frac{\<b_x\mu\>}{\<n\mu\>\s_{b_x n}}$, $\frac{\<1\mu\>}{\<n\mu\>\s_{1n}}$ only  depend on $b_x$ or $c_y$ respectively.

Though  $\Psi_{[1 a_2\cdots a_{i{-}\!1} n]}$ has particles with mixed  helicity, $h_{[1b_1b_2\cdots b_x]}$ and  $\tilde{h}_{[np_1p_2\cdots p_y]}$  do  have only particles of negative or positive helicity respectively.  Adding that closed cycles  vanish unless all of their elements have same helicity, $\mathrm{Pf'}{\bm\Psi}_n$  decouples to two parts which are dependent on particles of negative and positive helicity respectively,
\ba\label{rough}
\mathrm{Pf'}{\bm\Psi}_n=\left(\mathrm{sgn}(r)\sum_{\beta}h_{[1b_1\cdots b_x]}\sum_{I\cdots J}h_I\cdots h_J\right)\,\left(\mathrm{sgn}(\tilde{r})\sum_{\rho}\tilde {h}_{[1p_1\cdots p_y]}\sum_{K\cdots L}\tilde {h}_K\cdots \tilde {h}_L\right)\;.
\ea
Here we have explicitly written out the open cycles to emphasis them. $\beta,I,\cdots, J$ are the cycles of permutations $r$  of negative helicity particles except 1 and $\rho,K,\cdots,L$ are the cycles of permutations $\tilde{r}$ of positive helicity except $n$.

  For example, with $1^-2^-3^+4^+$,
 \ba
 \mathrm{Pf'}{\bm\Psi}_4=\Big(h_{[1]}h_{(2)}  +h_{[12]}\Big)\Big(\tilde{h}_{[4]}\tilde{h}_{(3)}+\tilde{h}_{[43]}\Big)\;,
  \ea
 with $1^-2^-3^-4^+5^+$,
 \ba
 \mathrm{Pf'}{\bm\Psi}_5=&&\Big(h_{[1]}h_{(2)}h_{(3)}+h_{[1]}h_{(23)}+
 h_{[12]}h_{(3)}+
h_{[132]}+ h_{[13]}h_{(2)}+
h_{[123]}
\Big)\nl
&& \times
 \Big(\tilde{h}_{[5]}\tilde{h}_{(4)}+\tilde{h}_{[54]}\Big)\;.
 \ea

Without the loss of generality, we let $-=\{1,2,\cdots,k\}$ and $+=\{k{+}1,k{+}2,\cdots,n\}$.
We try to prove the two parts in \eqref{rough} combine to two reduced determinants of matrices  ${\bm h}_{k}$, ${\bm {\tilde{h}}}_{n-k}$ respectively,  defined as $\mathrm{det'}\,{\bm h}_{k}=\frac{\mathrm{det}\,|{\bm h}_{k}|_{t_b}^{t_c}}{t_bt_c},\,\mathrm{det'}\,{\bm {\tilde{h}}}_{n-k}=\frac{\mathrm{det}\,|{\bm {\tilde{h}}}_{n-k}|_{t_p}^{t_q}}{t_pt_q}$ with $b,c\in -$,  $p,q\in +$ as ${\bm h}_{k}$ has a null vector $(t_1,t_2,\cdots,t_k)$ and ${\bm {\tilde{h}}}_{n-k}$ has a null vector $(t_{k+1},t_{k+2},\cdots,t_n)$.

Note that
 \ba
\mathrm{det}\;|\bm{h}_k|_1^{b_x}=(-)^{x}{\Big(\mathrm{det}\;|\bm{h}_k|_1^{1}\Big)\Big|}_{h_{b_xc}\rightarrow h_{1c}}
=(-)^{x}
\sum_{r\in S_{k-1}}{\rm sgn}{(r)}\,{ h_{(b_x\cdots)}\Big|}_{h_{b_xc}\rightarrow h_{1c}}h_{I}\cdots h_{J}\,.\qquad
 \ea
 Here $r$ is any permutation of particles of negative helicity except $1$, and $(b_x\cdots),I,\cdots,J$ are the cycles of $r$. $c$ can be anyone of $1,2,\cdots k$.

Since
\ba
h_{[1b_1b_2\cdots b_x]}&=&\frac{\<b_x\mu\>}{\<n\mu\>\s_{b_x n}}{h_{b_1b_2}\c\c h_{b_{x-1}b_x}h_{b_xb_1} \Big|}
_{h_{b_xb_1}\rightarrow h_{1b_1}}=\frac{\<b_x\mu\>}{\<n\mu\>\s_{b_x n}}{(b_1b_2\cdots b_x)_{_{h}} \Big|}\,,
_{h_{b_xb_1}\rightarrow h_{1b_1}}\qquad
\ea
 we write the first part in the RHS of \eqref{rough} as a sum over all possible  $b_x$, i.e. $b_x=1,2,\cdots k$ . This equality can also be seen by collecting terms with the same prefactor $\frac{\<b_x\mu\>}{\<n\mu\>\s_{b_x n}}$ ,
\ba
\mathrm{sgn}(r)\sum_{\beta}h_{[1b_1\cdots b_x]}\sum_{I\cdots J}h_I\cdots h_J
=\sum_{b_x=1}^k\left(\mathrm{sgn}(r)\sum_{\beta'}h_{[1\underbrace{b_1\cdots b_{x-\!1}}_{\beta'}b_x]}\sum_{I\cdots J}h_I\cdots h_J\right),\qquad
\ea
 Here $\beta'=\{b_1,\cdots,b_{x-1}\}$. $\beta',I,\cdots,J$ are the cycles of permutations of particles of negative helicity except $1$ and $b_x$. Then each term of the summation in RHS of the above equation equals  $\mathrm{det}\,|{\bm h}_k|_{1}^{b_x}$ up to a prefactor.
Summing over all possible $b_x$, i.e. $b_x=1,2,\cdots k$, gives the left parenthesis of RHS in \eqref{rough}. Similar derivations leads to the right parenthesis . Then
 \ba\label{11}
 \mathrm{Pf'}{\bm\Psi}_n=\Bigg(
 \sum_{b_x=1}^k\frac{\<b_x\mu\>}{\<n\mu\>\s_{b_x n}}
\mathrm{det}\,|{\bm h}_k|_1^{b_x}\Bigg)\Bigg(
\sum_{p_y=k+1}^n\frac{[\mu p_y]}{[1\mu]\s_{1p_y}}
\mathrm{det}\,|{\bm {\tilde{h}}}_{n-k}|_n^{p_y}\Bigg)\;.
\ea

We insert $\frac{t_nt_{b_x}}{t_1t_{b_x}}$ in every term of the first sum of above equation and $\frac{t_1t_{p_y}}{t_nt_{p_y}}$ in every term of the second sum, which doesn't change the value of $\mathrm{Pf'}{\bm\Psi}_n$. Then
\ba
 \mathrm{Pf'}{\bm\Psi}_n=\Bigg(
 \sum_{b_x=1}^k\frac{\<b_x\mu\>t_{b_x}t_ n}{\<n\mu\>\s_{b_xn}}
\frac{\mathrm{det}\,|{\bm h}_k|_1^{b_x}}{t_1t_{b_x}}
\Bigg)
 \;
\Bigg(
 \sum_{p_y=k+1}^n\frac{[\mu p_y]t_1t_{p_y}}{[1\mu]\s_{1p_y}}
\frac{\mathrm{det}\,|{\bm {\tilde{h}}}_{n-k}|_n^{p_y}}{t_nt_{p_y}}
\Bigg)\;.
 \ea
While all $\frac{\mathrm{det}\,|{\bm h}_k|_1^{b_x}}{t_1t_{b_x}}$  with $b_x=1,2,\cdots,k$ reduce to $\mathrm{det'}\,{\bm h}_k$, all
$\frac{\mathrm{det}\,|{\bm {\tilde{h}}}_{n-k}|_n^{p_y}}{t_nt_{p_y}}$ with $p_y=k+1,k+2,\cdots,n$ reduce to $\mathrm{det'}\,{\bm {\tilde{h}}}_{n-k}$. Then
\ba\label{combine}
 \mathrm{Pf'}{\bm\Psi}_n=\frac{\sum_{b_x=1}^{k}\frac{t_{b_x}t_n\<b_x\mu\>}{\s_{b_x n}}}{\<n\mu\>}\frac{\sum_{p_y=k+1}^n\frac{t_{1}t_{p_y}[\mu p_y]}{\s_{1p_y}}}{[1\mu]}
\mathrm{det'}\,{\bm h}_k
\mathrm{det'}\,{\bm {\tilde{h}}}_{n-k}\,.
 \ea
All gauge dependence of particle 1 and $n$ combine to one factor respectively and  on the support of 4d scattering equation \eqref{4dr},
 \begin{align}
t_n\sum_{b\in -'} {t_b\lambda_b^\alpha \over \s_{n\,b}}= \lambda_n^\alpha \,,
  \qquad
t_1 \sum_{p\in +'} {t_p\tilde\lambda^{\dot{\alpha}}_p \over \s_{1\,p}}=\tilde\lambda_1^{\dot{\alpha}}  \,,
\end{align}
the two prefactors before the determinants in \eqref{combine} reduce to $1$ respectively. Then we get
 \ba\label{detdet}
{\mathrm{Pf'}{\bm\Psi}_n\big|}_{k'=k}= \mathrm{det'}\,
{\bm h}_k\;\mathrm{det'}\,{\bm {\tilde{h}}}_{n-k}\;.
\ea
For example, with $1^-2^-3^-4^+5^+$,
 \ba
 \mathrm{Pf'}{\bm\Psi}_5&=&\Bigg(
 \Big(h_{(23)}+h_{(2)}h_{(3)}\Big) \frac{\<1\mu\>}{\<5\mu\>\s_{15}}
 + \Big({h_{(2)}\Big|}_{h_{22}\rightarrow h_{12}}
 h_{(3)}+{h_{(32)}\Big|}_{h_{23}\rightarrow h_{13}}\Big)\frac{\<2\mu\>}{\<5\mu\>\s_{25}}\nl
  &&\quad+\Big(h_{(2)}{h_{(3)}\Big|}_{h_{33}\rightarrow h_{13}}
  +{h_{(23)}\Big|}_{h_{32}\rightarrow h_{12}}\Big)\frac{\<3\mu\>}{\<5\mu\>\s_{35}}\Bigg)\nl
&&\times \Bigg(\frac{[\mu 5]}{[1\mu]\s_{15}}\tilde{h}_{(4)}+\frac{[\mu 4]}{[1\mu]\s_{14}}{\tilde{h}_{(4)}\Big|}_{\tilde{h}_{44}\rightarrow \tilde{h}_{54}}\Bigg) \nl
  &=&\sum_{a_x=1}^3\frac{\<b_x\mu\>}{\<5\mu\>\s_{b_x 5}}
\mathrm{det}\,|{\bm h}_3|_1^{a_x}
 \;\sum_{b_y=4}^5\frac{[\mu p_y]}{[1\mu]\s_{1p_y}}
\mathrm{det}\,|{\bm {\tilde{h}}}_{2}|_5^{b_y}\nl
&=&\mathrm{det}'\,{\bm h}_3\,\mathrm{det}'\,{\bm {\tilde{h}}}_2
 \ea
\section{Extension to all solution sectors}
We have arrived at  \eqref{11} without using the explicitly form of  1-length cycle, {\it i.e.} $C_{aa}$. When extended to all solution sectors, those cycles whose length are longer than 1 don't change, while the 1-length cycles change to $C_{aa} $ with the solutions of $k'$ sectors plugged in . That is, we need to  enhance the origin two matrices to ${\bm h}_k^{k'}$ and  ${\bm {\tilde h}}_{n-k}^{k'}$ with their diagonal entries  depending  on  the solution sector $k'$ while the off-diagonal entries unchanged. The expression of $C_{aa}$ with $a\in -$ and $a\in -'$ has been given in  \eqref{a22}. Note that this expression is true even when $k'\neq k$,
 \ba\label{haa1}
h_{aa}=-\sum_{\tiny{\substack{b\neq a\\b\in -'}}}\frac{t_b}{t_a}\frac{\<ab\>}{\s_{ab}}\quad a\in- \  {\text {and}}\ a\in-'\,.
\ea
Now we derive the expression of $C_{aa}$ with $a$ not consistent in helicity sector and solution sector.  When $a\in -$ but  $a\notin -'$, we have
\ba
C_{aa}=&&-\sum_{{p\in {\rm +'},\,p\neq a}}\frac{\<ap\>[p\mu]}{[a\mu]\s_{ap}}
-\sum_{b\in {\rm -'}}\frac{\<ab\>[b\mu]}{[a\mu]\s_{ab}}
\ea
After we plug in the changes of variables \eqref{changes}, unlike \eqref{a21}, terms with $a\in +'$ and $p=a$  both contribute.
\ba
C_{aa}=&&-\frac{1}{[a\mu]}
\sum_{\substack{p\in {\rm +'},\,p\neq a\\b\in {\rm -'}}}
\Big(\frac{\<a b\>t_bt_p[p\mu]}{\s_{pb}\s_{ap}}+\frac{\<a b\>t_bt_p[p\mu]}{\s_{bp}\s_{ab}}
\Big)+\sum_{b\in {\rm -'}}\frac{\<ab\>t_bt_a}{\s_{ba}^2}
\ea
The first term on the RHS also factorizes into two parts following the trick used in \eqref{a21}, \eqref{a22},
\ba
-\frac{1}{[a\mu]}
\sum_{\substack{p\in {\rm +'},\,p\neq a\\b\in {\rm -'}}}
\frac{\<a b\>t_bt_p[p\mu]}{\s_{ab}\s_{ap}}=-\left(\sum_{{b\in {\rm -'}}}
\frac{t_at_b\<a b\>}{\s_{ab}}\right)\left(
\sum_{p\in {\rm +'},\,p\neq a}\frac{t_p[p\mu]}{t_a\s_{ap}[a\mu]}\right)=0\,,
\ea
while it vanishes as shown in the last equality because  the part in the first  parenthesis vanishes on the support of 4d scattering equation \eqref{4dr}, (note that $a\in +'$)
\begin{align}
t_a\sum_{b\in -'} {t_b\lambda_b^\alpha \over \s_{a\,b}}=\lambda_a^\alpha
\end{align}
then we see that $C_{aa}$ only has contribution from the term of $p=a$, and we obtain
\ba
C_{aa}
=-\sum_{\substack{b< c\\b,c\in {\rm -'}}}\frac{\<bc\>t_bt_ct_a^2\s_{bc}}{\s_{ba}^2\s_{ca}^2}\qquad {\rm when}\ a\in -\; \text{but}\; a\notin {\rm -'}\;.
\ea
Consequently, we have
 \ba\label{haakp}
{h}_{aa}&=&-t_a^2\sum\limits_{b<c;\ b,c\in {\rm -'}}\frac{t_b t_c \s_{bc}\<bc\>}{\s_{ab}^2\s_{ac}^2}\quad a\in -\; \text{but}\; a\notin {\rm -'}\;,
\ea
By a parity transformation, we can directly obtain $\tilde{h}_{aa}$
\ba\label{haakp2}
\tilde{h}_{aa}&=&-\sum_{\tiny{\substack{b\neq a\\b\in +'}}}\frac{t_b}{t_a}\frac{[ab]}{\s_{ab}}\qquad\qquad\qquad a\in+ \  {\text {and}}\ a\in+'\nl
\tilde{{h}}_{aa}&=&-t_a^2\sum\limits_{b<c;\ b,c\in {\rm +'}}\frac{t_b t_c \s_{bc}[bc]}{\s_{ab}^2\s_{ac}^2}\quad a\in +\; \text{but}\; a\notin {\rm +'}\;.
\ea

When $k'=k$, these extended matrices come back to their original ones. When $k'\neq k$, one of
 $\mathrm{det}\,{\bm h}_k^{k}$, $\mathrm{det}\,{\bm {\tilde{h}}}_{n-k}^{k}$ must vanish. Further more, when $k'<k$, after deleting appropriate row and column of ${\bm h}_k^{k}$, the determinant of the remaining matrix still vanishes, so does the ${\bm {\tilde{h}}}_{n-k}^{k}$ when $k'>k$, which results in the vanishing of $\mathrm{Pf'}{\bm\Psi}_n$ in $k'\neq k$ sectors. We will discuss this in sec.\ref{s41}.   Some integrands  receive the contribution from  the $k'\neq k$ sectors, such as ${\cal P}_n$, which will be discussed in sec.\ref{s42}.

\subsection{the vanishing of reduced Pfaffian in other sectors}\label{s41}
 We start from the equation \eqref{11}. Note that we have got this by deleting the $1^{\rm th}$ and $n^{\rm th}$ rows and columns of ${\rm \Psi}_n$. We can also delete other rows and columns to get a similar expression. What's more, along the demonstration of \eqref{pin} to \eqref{11}, we don't use the scattering equations \eqref{4dr}, in other words \eqref{11} is true for any solutions. After \eqref{11}
  , the scattering equation is used and we  demonstrate \eqref{detdet}. Now we move to other solution sectors. Without loss of generality, let's consider $-=\{1,2,\ldots, k\}$ while $-'=\{1,2,\ldots,k'\}$ , which makes our discussion simpler. Then when $k'<k$, further on,  we can also and always delete $(k'+1)^{\rm th}$ and $n^{\rm th}$ column and row instead of the of $1^{\rm th}$ and $n^{\rm th}$ ones, then the reduced Pfaffian becomes
 \ba\label{kp1p}
 \mathrm{Pf'}{\bm\Psi}_n&=& \Bigg(
 \sum_{b_x=1}^k\frac{\<b_x\mu\>}{\<n\mu\>\s_{b_x n}}
\mathrm{det}\,|{\bm h}_k^{k'}|_{k'+1}^{b_x}
\Bigg)
 \;
 \Bigg(
 \sum_{p_y=k{+}1}^n\frac{[\mu p_y]}{[k'{+}1,\mu]\s_{k'{+}1,p_y}}
\mathrm{det}\,|{\bm {\tilde{h}}}_{n-k}^{n-k'}|_n^{p_y}
\Bigg)
\qquad
\ea
Notice that we still have to  calculate the determinants of series of matrices. Instead of both summations in RHS of  \eqref{kp1p} being combined to  simple factors as shown in \eqref{combine}, we  show that the determinants of matrices $|{\bm h}_k^{k'}|_{k'+1}^{b_x}$ with $b_x=1,2,\cdots,k$ in the first summation vanish identically.


These matrices all come from the original matrix ${\bm h}_k^{k'}$ with the $(k'+1)^{\rm th}$ column deleted and the
 $1^{\rm th},2^{\rm th},\cdots,k^{\rm th}$
 row deleted respectively.  An important observation is that the first $k'$ columns of these matrices are  linearly dependent as
 \ba\label{lineardependent}
 t_1{\bm\eta}_1+t_2{\bm\eta}_2+\cdots t_{k'}{\bm\eta}_{k'}={\bm 0}
 \ea
 here ${\bm\eta}_1,{\bm\eta}_2,\cdots,{\bm\eta}_{k'}$ are the $1^{\rm th},2^{\rm th},\cdots,{k'}^{\rm th}$ columns of anyone of matrices $|{\bm h}_k^{k'}|_{k'+1}^{b_x}$ with $b_x=1,2,\cdots,k$. This is equivalent to say that
 \ba\label{tbhab}
 \sum_{b=1}^{k'}t_bh_{ab}=0,\qquad\quad \text{for} \ a=1,2,\cdots,k
 \ea
 These equations come from two totally different origins as $a\leq k'$ or $a>k'$.

 For the case of $a\in-'$ {\it i.e.} $a=1,2,\cdots,k'$, the establishment of \eqref{tbhab} come from the fact that
 the diagonal elements $h_{aa}$ are a linear combination of some off-diagonal entries as shown in \eqref{haa1} with some appropriate coefficients.

 While for the cases of $a>k'$, note that $a$ now belongs to the set $+'$,  and  the validity of \eqref{tbhab} come from the change of variables~\eqref{changes}. What we need here is the cases of $a=k'+1,k'+2,\cdots,k$,
\begin{align}
t_a\sum_{b\in -'} {t_b\lambda_b^\alpha \over \s_{a\,b}}= \lambda_a^\alpha \qquad \text{for} \ a=k'+1,k'+2,\cdots,k
\end{align}
Obviously after we act $\lambda_{a}$  on both sides of above equation , both sides vanish, that is
 \ba
\sum_{b=1}^{k'}t_bh_{ab}=\sum_{b=1}^{k'}t_b\frac{\<ab\>}{\s_{ab}}=0\qquad \text{for} \ a=k'+1,k'+2,\cdots,k
 \ea

After understanding \eqref{lineardependent}, now it is easy to understand the vanishing of all   matrices $|{\bm h}_k^{k'}|_{k'+1}^{b_x}$ with $b_x=1,2,\cdots,k$.
 We  take a multiple of the $a^{\rm th}$ row  of these matrice by $t_a$  for $a=1,2,\cdots ,k'$, and then add $2^{\rm th},3^{\rm th},\cdots, k'^{\,{\rm th}}$ column to the $1^{\rm th}$ column; in this way we obtain a new $1^{\rm th}$  column whose entries all equal to zero because of  \eqref{lineardependent}. Since we just do some fundemantal operation on these matrices and we obtain a column with all entries equal to zero, all determinants of these matrices vanish.

 When $k'<k$, the determinants of matrices  in the first summation in \eqref{kp1p} vanish; when $k'>k$, the determinants of matrices  in the second summation in \eqref{kp1p} vanish.
 $\mathrm{Pf'}{\bm\Psi}_n$ only receives the contribution of $k'=k$ sector, then we proved the identity \eqref{id1} given in the introduction.

In the reduction of $\mathrm{Pf'}{\bm\Psi}_n$, we reorganize the Pfaffian using some fundamental (almost) gauge invariant objects and then deal with these objects , finally we reconstruct the reduced Pfaffian using $\det {\bm h}_k$ and $\det {\bm {\tilde h}}_{n-k}$. This procedure is quite general. The reduced Pfaffian can be thought as putting two kinematic of the deleted particles in higher dimensions and the Lorentz contraction of them and anything else vanish unless contraction of them each other equal to 1 ~\footnote{Here is some subtlety. We  have to complex $k_1,k_n$ set in higher dimensions such that they dotting anything else equal to 0, while they dotting each other equal to 1. It is just a mathematics trick after all nothing about  $k_1,k_n$ changes beyond the reduced Pfaffian. }. Similarly, we can put polarizations of $m$ pairs of particles in higher dimension and then we get the reduced compactified Pfaffian, which is the integrand of Einstein-Maxwell, Einstein-Maxwell-scalar, Yang-Mills-scalar, Born-Infeld amplitudes. This can be viewed as the ``fancy reduced Pfaffian" as we may meet several open brackets in each term of the expansion. Further more, the reduced squeezed Pfaffian, which is the integrand of Einstein-Yang-Mills, can be obtained by some combination of the reduced compactified Pfaffian.  So all these methods can be applied in the reduced squeezed Pfaffian,too.These are presented in the Appdendix~\ref{compactified},\ref{squeezed}.

 There are some integrands that can't be organized as a matrix, let alone its Pfaffian, such as the ${\cal P}_n$ used in $F^3,R^2,R^3$ amplitudes. We can still reduce it into objects related to $\det {\bm h}_k$ and $\det {\bm {\tilde h}}_{n-k}$ and make some properties manifest.  Besides, it receives the contribution from several sectors and one need to add up all of these to get the corresponding amplitudes, as discussed in the following subsection.

\subsection{the new object ${\cal P}_n$ for higher dimension operator}\label{s42}
 As shown in~\cite{He:2016iqi}, as a generalization of the reduced Pfaffian in Yang-Mills theory, ${\cal P}_n$ is a new, gauge-invariant object that leads to gluon amplitudes with a single insertion of $F^3$, and gravity amplitudes by Kawai-Lewellen-Tye relations. When reduced to four dimensions for given helicities, this new object vanishes for any solution of scattering equations on which the reduced Pfaffian is non-vanishing. This intriguing behavior in four dimensions explains the vanishing of graviton helicity amplitudes produced by the Gauss-Bonnet $R^2$ term, and provides a scattering-equation origin of the decomposition into self-dual and anti-self-dual parts for $F^3$ and $R^3$ amplitudes.

 No matter what ${\cal P}_n$ is, it must be gauge invariant. It's most natural to start from the expansion of Pfaffian in a manifest gauge invariant way \eqref{psi1}.
Reorganize these gauge invariant objects according to their length and we define the minimal gauge invariant and guage invariant objects
 $P$ as
\ba\label{P}
P_{i_1\,i_2\,\cdots\,i_r}
:=\sum_{\tiny{\substack{|I_1|=i_1,|I_2|=i_2,\cdots,|I_r|=i_r}}}\Psi_{I_1}\Psi_{I_2}\cdots\Psi_{I_r}\;,
\ea
with $ i_1+ i_2+\cdots +i_r=n$  and the convention $i_1\leq i_2\cdots \leq i_r $.

Then Pfaffian can be written as
\ba\label{cp}
{\rm Pf}\bm{\Psi}_n=\sum_{i}
\substack{(-)^{n-m}}P_{i_1i_2\cdots i_m}=0\;,
\ea

Decorated with some appropriate coefficient, $P$ can be used to build up some unknown CHY integrands, such as the new integrand is defined as
\ba\label{2}
\mathcal{P}_n=\displaystyle \sum\limits_{\tiny \substack{1\leq i_1\leq i_2\leq\cdots\leq i_m\leq n \\i_1+i_2+\cdots+i_m=n}}~(-)^{n-m}~\left( N_{i>1} + c\right)~P_{i_1\,i_2\,\cdots\,i_m}\;,
\ea
where $N_{i>1}$ denotes the number of indices in $i_1, i_2, \cdots, i_m$ which are larger than $1$, or the number of cycles with length at least $2$; $c$ is just any constant because we can add any multiplet of \eqref{cp} without changing the answer.

In 4d, ${\rm Pf}\bm{\Psi}_n$ reduce to the determinants of ${\bm h}_k^{k'}$ and ${\bm {\tilde{h}}}_{n-k}^{k'}$. Similarly we define
\ba\label{H}
H_{i_1i_2\cdots i_{\ell}}
=\sum_{\tiny{\substack{|I_1|=i_1,|I_2|=i_2,\cdots,|I_t|=i_{\ell}}}}h_{I_1}h_{I_2}\cdots h_{I_{\ell}}\;,
\ea
with $ i_1+ i_2+\cdots +i_{\ell}=k$  and the convention $i_1\leq i_2\cdots \leq i_{\ell} $. Then $\mathrm{det}\;\bm{h}_k^{k'}$ can be rewritten as a sum of $H$ and similarly works $\mathrm{det}\; \bm{\tilde{h}}_{n-k}^{k'}$,
\ba\label{Hdet}
\mathrm{det}\;\bm{h}_k^{k'}=\sum_{\{i\}_k^{\ell}}(-)^{k-{\ell}}H_{i_1i_2\cdots i_{\ell}}\;.
\qquad
\mathrm{det}\; \bm{\tilde{h}}_{n-k}^{k'}=\sum_{\{{\tilde i}\}_{n-k}^{{\tilde \ell}}}(-)^{n-k-{{\tilde \ell}}}\tilde{H}_{{\tilde i}_1{\tilde i}_2\cdots {\tilde i}_{{\tilde \ell}}}\;,
\ea
where we have introduced shorthand notation for the summation range,  $\{i\}_k^{\ell}$ means $i_1+ i_2 + \ldots i_{\ell}=k$ and $i_1 \leq i_2 \leq \cdots \leq i_{\ell}$ and similarly for $\{{\tilde i}\}_{n-k}^{{\tilde \ell}}$.

Further, similar to the definition of \eqref{2}, we define
two auxiliary objects $\mathcal{H}_k^{k'}$ and $\mathcal{\tilde{H}}_{n-k}^{k'}$,
\ba
\mathcal{H}_k^{k'}=\sum_{\{i\}_k^{\ell}}(-)^{k-{\ell}}{\substack{N_{i>1}}}H_{i_1i_2\cdots i_{\ell}}\;, \quad \mathcal{\tilde{H}}_{n-k}^{k'}=\sum_{\{\tilde{i}\}_{n-k}^{\tilde{{\ell}}}}(-)^{n-k-\tilde{{\ell}}}
{\substack{N_{\tilde{i}>1}}}\tilde{H}_{\tilde{i}_1\tilde{i}_2\cdots \tilde{i}_{\tilde{{\ell}}}}\;.
\ea
Thanks to \eqref{psiI}, $P$ reduces to  several products $H$ and $\tilde{H}$ in 4d ,
\ba\label{ph}
P_{i_1i_2\cdots i_m}= \sum_{j,\tilde{j}} H_{j_1j_2\cdots j_{\ell}}\tilde{H}_{\tilde{j}_1\tilde{j}_2\cdots\tilde{j}_{\tilde{{\ell}}}}\;,
\ea
Here the sum is over all distinct partition of $i_1i_2\cdots i_m$ into two parts $j_1j_2\cdots j_{\ell}$ and $\tilde{j}_1\tilde{j}_2\cdots\tilde{j}_{\tilde{{\ell}}}$, with $j_1+ j_2+\cdots+ j_{\ell}=k$ and $\tilde{j}_1+\tilde{j}_2+\cdots+\tilde{j}_{\tilde{{\ell}}}=n-k$.
Dividing $N_{i>1}$ in \eqref{2} into two parts $N_{j>1}$ and $N_{\tilde{j}>1}$ (set $c=0$) which depend on $-$ and $+$ sets respectively,  ${\cal P}_n$ reduces to:
\ba
\mathcal{P}_n
=
&&\left(
\sum_{\{i\}_k^{\ell}}(-)^{k-{\ell}}{\substack{N_{i>1}}}H_{i_1i_2\cdots i_{\ell}}
\right)
\left(
\sum_{ \{\tilde{i}\}_{n-k}^{\tilde{{\ell}}}}(-)^{n-k-\tilde{{\ell}}}\tilde{H}_{\tilde{i}_1\tilde{i}_2\cdots \tilde{i}_{\tilde{{\ell}}}}
\right)\nl
&&+
\left(
\sum_{\{i\}_k^{\ell}}(-)^{k-{\ell}}H_{i_1i_2\cdots i_{\ell}}
\right)
\left(
\sum_{ \{\tilde{i}\}_{n-k}^{\tilde{{\ell}}}}(-)^{n-k-\tilde{{\ell}}}{\substack{N_{\tilde{i}>1}}}
\tilde{H}_{\tilde{i}_1\tilde{i}_2\cdots \tilde{i}_{\tilde{{\ell}}}}
\right)
\ea
where each mix summation over $i$ and $\tilde{i}$ decouples to two independent summation over $i$ and
 over $\tilde{i}$ respectively. Then
\ba\label{p1}
{\mathcal{P}_n\big|}_{k'}
= \mathcal{H}_k^{k'}\;\mathrm{det}\,{\bm{\tilde h}}_{n-k}^{k'}\;+\mathrm{det}\,{\bm {h}}_{k}^{k'}\;\mathcal{\tilde{H}}_{n-k}^{k'}
\;.
\ea
When $k'=k$, both terms in the RHS of \eqref{p1} vanish, which is orthogonal to ${\rm Pf'}\bm{\Psi}_n$ and answers the vanishing of $R^2$ theory which is a Gauss-Bonnet term in $4$ dimensions. When $k'<k$, the second term  vanishes and   $\mathcal{P}_n$ reduces to $\mathcal{H}_k\;\mathrm{det}\,{\bm{\tilde h}}_{n-k}^{k'}$, which gives the self-dual amplitude of $F^3,R^3$ theory {\it etc.} When $k'>k$, the first term  vanishes and   $\mathcal{P}_n$ reduces to $\mathrm{det}\,{\bm {h}}_{k}^{k'}\;\mathcal{\tilde{H}}_{n-k}$, which gives the anti-self-dual amplitude of those theory.

For example, the integrand for $F^3$ theory reads ${\cal I}_n^{ F^3}={\cal C}_n~\mathcal{P}_n$. In 4 dimension, when $k'<k$, $I_{n}^{F^3}$ reads
\ba
I_{n}^{F^3}=\frac{{\cal C}_n~\mathcal{H}_k^{k'}\;\mathrm{det}\,{\bm{\tilde h}}_{n-k}^{k'}}{\mathrm{det'}\,{\bm{h}}_{k'}\mathrm{det'}\,{\bm{\tilde{h}}}_{n-k'}}
\ea
Here $\mathrm{det'}{\bm{h}}_{k'}\mathrm{det'}\,{\bm{\tilde{h}}}_{n-k'}$ comes from the transition of two forms of scattering equations as shown in \eqref{id2}.

\section{Discussion}

In CHY representation, the fundamental gauge invariant objects are quite common, either $C_{aa}$ or the trace of some linearised field strength together with some $\s$'s.  In this paper, we find a rather general way to reduce this gauge invariant objects into that made up of spinors using 4d scattering equations. Particularly, we show how the reduced Pfaffian reduces to some determinants and why it vanishes on the support of most solutions. This explains why only some particular solutions contribute to the YM or GR amplitudes  according to their helicity structure in 4d and provides a basis for dividing the solutions of scattering equations into MHV,NMHV,$\cdots$,$\overline{\rm MHV}$ sectors which contributes to corresponding YM or GR amplitudes respectively, also seen in \cite{Du:2016fwe}.

 We extend this methods to reduced compactified Pfaffian where some polarizations are set in higher dimensions, which is building block for EM, EMS, YMS or DBI  amplitudes. We give the explicit reduction results of the compactified Pfaffian up to 3 pairs of particles whose polarizations are set in higher dimension and provides the general way to get the reduction with arbitrary pairs.  Another interesting integrand with polarization involved is the reduced squeezed Pfaffian, which is the building block of EYM theory \cite{Cachazo:2014nsa,Cachazo:2014xea,Nandan:2016pya,Adamo:2015gia}. As it can be expressed by some combination of the reduced compactified Pfaffian, its reduction in 4d is directly obtained from that of the reduced compactified Pfaffian. The results of one and two gluon color traces are explicitly presented in the Appendix \ref{squeezed}.

 Even some integrands which can't be organized as a matrix, let alone its Pfaffian,  such as the new integrand ${\cal P}_n$ used in $F^3,R^2,R^3$ theory, also can be enclosed in this procedure. We decompose ${\cal P}_n$  to some fundamental gauge invariant objects,reduce these fundamental gauge invariant objects first and then  organize them into a compacted form, which apparently shows most information of the ${\cal P}_n$, and explains the orthogonality of $F^3$ and YM amplitudes, the vanishing of Gauss-Bonnet term  $R^2$ and the self-dual and anti-self-dual amplitudes of $F^3$ or $R^3$ amplitudes in 4d. In fact,  we use these properties to guess  what the compacted form in 4d of ${\cal P}_n$ should be, then fix the coefficient of  the fundamental gauge invariant objects and finally get the ${\cal P}_n$. This is quite general to find the CHY¡¡integrand of an unknown theory. Even when the scattering equations  or $\s$ dependence of the entries in the matrix ${\bm \Psi}_n$ has been changed, their CHY integrand are very likely to be decomposed into some $C_{aa}$ or trace like fundamental gauge invariant objects. And we can reduce these objects first, organize them into a form manifest showing some properties the theory requires and finally confirm their CHY integrand. This can even be applied at loop level, as shown in \cite{He:2016mzd,He:2017spx}.

 Instead of reducing all kinds of integrands in 4d, we now turn to the general 4d CHY formulae.
 After overcoming the obstacles to reduce integrands with polarization involved, the calculation of CHY formulae becomes much simpler. We develop the 4d CHY formulae to directly calculate the amplitude of the some theory. The reduced Pfaffian behaves like a solution filter, making the building of 4d CHY formulae natural. As if the general CHY formulae has been reduced to 4d CHY formulae and the number of  solutions decrease from $(n-3)!$ to $E_{n{-}3,k'{-}2}$.

  We have discussed the reduction of the reduced compactified Pfaffian and squeezed Pfaffian in Appendix \ref{compactified},\ref{squeezed} and discussed how the valid solution sector shift from the helicity sector. The more polarizations there are , the more efficient our procedure is. Even when there is no polarization involved, and the reduced compactified Pfaffian totally  reduce to ${\rm Pf}'{\bm A}_n$ times something, our reduction procedure still holds and  it tells us only the $k'=n/2$ solution sector contributes. This means CHY formulae of some effective field theory such as Born-Infeld, Dirac-Born-Infeld, Non-Linear Sigma Model, Special Galileon theories with ${\rm Pf}'{\bm A}_n$  acting as CHY integrand also reduce to a set of 4d CHY formulae. Even for some theories that receive the contribution of several solution sectors such as those with ${\cal P}_n$ acting as CHY integrand, the physical meaning of 4d CHY formulae is also  apparent: the contribution from the $k'<k$ sectors gives self-dual amplitude and that of $k'>k$ gives anti-self-dual amplitudes.

Many good properties shared by CHY formulae are still inherited by the 4d CHY formulae.  The soft limits has been discussed in~\cite{He:2016vfi,Cachazo:2015ksa,He:2014bga,Bianchi:2014gla,Schwab:2014xua}. Factorisation should also be easy to study. Not only the CHY formulae have a simple representation in 4d, 4d CHY formulae can also help us understand the CHY formulae in general dimension. Besides, the supersymmetrization of the 4d CHY formulae is directly and we just need to replace the $\tilde\lambda_a^\a$ with $\{\tilde\lambda_a^\a|\eta_a^A\}$ in the scattering equation in \eqref{int2} as shown in \cite{He:2016vfi}. This way we can involve fermions in CHY formulae, for example we use SYM amplitudes to build up QCD amplitudes as shown \cite{He:2016dol}. In the same paper, we use two set of spinors to describe the massive Higgs,  which has been   generalized to calculate form factors~\cite{He:2016jdg,Brandhuber:2016xue} .

We tentatively study whether the solutions divide  beyond in 4d .  Especially we hope something interesting come out in 6 dimensions where we also have a good spinor representation  \cite{Bern:2010qa,Brandhuber:2010mm,Dennen:2010dh}  and some nice result of CHY formulae in 6d comes out. We can treat a massive particle in 4d as a massless particles in 6d, especial the massive loop particles in 4d. Up to now, our result is negative and we didn't find the solutions of scattering equation divide again in other dimension.

CHY formulae has been extended to 1-loop level, as discussed in~\cite{He:2015yua,Cachazo:2015aol}. It has been known that what underpins the CHY formulae is ambitwistor string.  And ambitwistor string theory has been extended to 1-loop level, as shown in ~\cite{Adamo:2013tsa,Geyer:2015jch,Geyer:2015bja}. However we find the solutions at 1-loop level don't  divide into several sectors again in SYM or SUGRA  theory.  CHY formulae has singular solutions at 1-loop,  how about 4d CHY formulae and how does it contribute to the divergence bubble or tadpole? Also  it is interesting to check whether the integrand  still factorizes to two objects that depend on particles of negative helicity or positive helicity respectively. Also it will help us to build the general 1-loop CHY integrand.

As discussed in~\cite{Cachazo:2016ror}, also it is very useful to study  the positivity  of the Jacobian or integrand in 4d CHY formulae. \eqref{iden2} is a useful identity  as it link several objects.  As shown in ~\cite{Cachazo:2016ror}, $\det{}' \Phi_n (\{s_{a b}, \sigma_a\})$ is positive at the positive region. We know that $~\mathrm{det'}\,{\bm{h}}_{k'}\mathrm{det'}\,{\bm{\tilde{h}}}_{n-k'}$  is exactly the result of ${\rm Pf}'{\bm {\Psi}}_n$ with $k'$ external particles of negative helicity. If the 4d Jacobian $J_{n,k'}$ is also positive at the positive region, it strongly supports that the YM amplitude is also positive in some regions.

\section*{Acknowledgments}
Yong Zhang thanks S. He and F. Teng very much for discussions. Y.Z.'s research is partly supported by NSFC Grants No. 11235003.

\appendix

\section{A conjecture on $J_{n,k'}$ }\label{jnk}

In this section, without making confusion, we denote $\frac{\s_{a}-\s_{b}}{t_at_b}=:(ab)$ . We have strong evidence to support the following conjecture,

\ba
J_{n,k'}=\frac{1}{\prod\limits_{a=1}^nt_a^2\prod\limits_{{b\in -',p\in+'}}(bp)^2}\sum\limits_{b{<}c,p{<}q}\left(
 \prod\limits_{\substack {1{\leq} x{\leq} n{-}k'{-}2\\1{\leq} y{\leq} k'{-}2}}(b_xc_x)\<b_xc_x\>(p_yq_y)[p_yq_y](d_{xy}r_{yx})^2\right)\quad
\ea

Here the sum is over all $b_1{<}c_1,b_2{<}c_2,\cdots,b_{n-k'{-}2}{<}c_{n-k'{-}2};\;b_1,c_1,b_2,c_2,\cdots,b_{n-k'{-}2},c_{n-k'{-}2}\in -'$
and $p_1{<}q_1,p_2{<}q_2,\cdots,p_{k'{-}2}{<}q_{k'{-}2};$$\;p_1,q_1,p_2,q_2,\cdots,p_{k'{-}2},q_{k'{-}2}\in +'$. The $x^{\rm th}$ row of matrix $d$ is $-'\backslash\{b_x,c_x\}$ for $1{\leq} x{\leq} n{-}k'{-}2$ and the $y^{\rm th}$ row of matrix $r$ is $+'\backslash\{p_y,q_y\}$ for $1{\leq} y{\leq} k'{-}2$. When $k'{=}2$ or $n-2$, $(d_{xy}r_{yx})^2$ reduces to 1. We have numerically checked this formula up to 9 points in all solution sectors, and 15 points in $k'=3$ sector.

Here are some examples.

For the MHV solution sector, it can be analytically proved  that
\ba\label{jn2}
J_{n,2}=\frac{(bc)^{n-4}\<bc\>^{n-4}}{\prod\limits_{a=1}^nt_a^2\prod\limits_{{p\in+'}}(bp)^2(cp)^2}
\ea
Here $\{b,c\}=-'$. After deleting the 4 columns and rows about $b,c$ of the matrix in \eqref{J} with a compensate $\frac{\prod_{a=1}^n t_a}{t_b^3t_c^3(bc)^2\<bc\>^2}$, the minor becomes a ``diagonal" matrix whose diagonal entries are $2\times 2$ matrices and their determinants can be easily calculated out as $\frac{(bc)\<bc\>}{(pb)^2(pc)^2t_p^3}$ for $p=+'$, then we get \eqref{jn2}.

For the NMHV solution sector,
\ba
J_{6,3}=\frac{1}{\prod\limits_{a=1}^nt_a^2\prod\limits_{{b\in -',p\in+'}}(bp)^2}\sum\limits_{b{<}c,p{<}q}
 (bc)\<bc\>(pq)[pq](dr)^2\quad
\ea
Here $d,r$ is a particle label as the breviate of $d_{11},r_{11}$  and $\{b,c,d\}=-' $, $\{p,q,r\}=+'$.
\ba
J_{7,3}=\frac{1}{\prod\limits_{a=1}^nt_a^2\prod\limits_{{b\in -',p\in+'}}(bp)^2}\sum\limits_{b_1{<}c_1,b_2{<}c_2,p{<}q}
 (b_1c_1)\<b_1c_1\>(b_2c_2)\<b_2c_2\>(pq)[pq](d_{1}r_{1})^2(d_{2}r_{2})^2\qquad
\ea
Here $d_{1},d_{2}$ is a particle label as the breviate of $d_{11},d_{21}$ and $\{b_1,c_1,d_{11}\}=-' $,$\{b_2,c_2,d_{21}\}=-' $.
 $r_{1},r_{2}$ is a particle label as the breviate of $r_{11},r_{12}$ and $\{p,q,r_{11},r_{12}\}=+'$. Note that this restrain doesn't fix $r_{11},r_{12}$ totally as $r_{11},r_{12}$ can exchange their value. However it doesn't affect the value of $J_{7,3}$ as we always sum over all $b_1<c_1,b_2<c_2$.
\ba
J_{n,3}=\frac{1}{\prod\limits_{a=1}^nt_a^2\prod\limits_{{b\in -',p\in+'}}(bp)^2}\sum\limits_{b{<}c,p{<}q}\left(
 \prod\limits_{\substack {1{\leq} x{\leq} n{-}5}}(b_xc_x)\<b_xc_x\>(pq)[pq](d_{x}r_{x})^2\right)\quad
\ea
Here $d_{x}$ is a particle label as the breviate of $d_{x1}$ and $\{b_x,c_x,d_x\}=-' $.
 $r_{x}$ is a particle label as the breviate of $r_{1x}$ and $\{r_{11},r_{12},\cdots r_{1,n-5}\}=+'\backslash\{p,q\} $.

For NNMHV solution sector,
\begin{align}
J_{8,4}=\frac{1}{\prod\limits_{a=1}^nt_a^2\prod\limits_{{b\in -',p\in+'}}(bp)^2}
 \sum\limits_{\substack {b_1{<}c_1,b_2{<}c_2\\p_1{<}q_1,p_2{<}q_2}}
& (b_1c_1)\<b_1c_1\>(b_2c_2)\<b_2c_2\>(p_1q_1)[p_1q_1](p_2q_2)[p_2q_2]\nl
&\times (d_{11}r_{11})^2(d_{21}r_{12})^2
 (d_{21}r_{12})^2(d_{22}r_{22})^2\,.
\end{align}
Here  $\{b_1,c_1,d_{11},d_{12}\}=-' $, $\{b_2,c_2,d_{21},d_{22}\}=-' $,
$\{p_1,q_1,r_{11},r_{12}\}=+'$,$\{p_2,q_2,r_{21},r_{22}\}=+'$. Note that this restrain doesn't fix $d_{11},d_{12}$ totally as $d_{11},d_{12}$ can exchange their value neither does $r_{11},r_{12}$. In most cases, it doesn't affect the value of $J_{8,4}$ however a few cases do rely on particular ordering of $d_{11},d_{12}$ or $r_{11},r_{12}$, leaving a further study to fix the final expression of $J_{n,k'}$.

\section{The reduced compactified Pfaffian in 4d}\label{compactified}
In the main text, we have discussed the reduced Pfaffian which we delete $1^{\rm th}$ and $n^{\rm th}$ rows and columns of the matrix ${\bm \Psi}_n$ . Then we introduce the open cycle to reduce it into two reduced determinants. Also we can effectively think that we set the momenta of the particles $1,n$ in higher dimension and they dotting everything equal to zero unless they dotting themselves equal to 1 to make the complement $\frac{1}{\s_{1n}}$. Then we can still decompose the reduced Pfaffian into some (modified) closed cycles.
  One closed cycle must contain $1,n$ both or vanish if it just contains one of them, and then it reduces to open cycle as $k_1,k_n$ are set in higher dimension.  This can be extended to other cases as now we set the polarisation of some pairs of particles in higher dimension. We call it reduced compactified Pfaffian which is the building block for EM, EMS, YMS  amplitudes, as discussed in ~\cite{Cachazo:2014xea}. Then we can use the similar trick to reduce the reduced compactified Pfaffian in 4d.

We denote the set of particles whose polarisation are set in higher dimension as $\upgamma$ and those that are not as ${\texttt h}$. Besides, we divide ${\texttt h}$ into ${\texttt h}^-$ and ${\texttt h}^+$ whose helicity are negative and positive respectively and denote $\tilde k= |{\texttt h}^-|$. Obviously, the length of set $\upgamma$ must be even. Further on, we let the polarisation of particles in $\upgamma$ be anyone of an orthogonal bases and they dotting each other equal to 1 or 0, denoted as $\d^{I_aI_b}$.  Then the compactified Pfaffian can be think of as being deleted the rows and columns of $\upgamma$ in the last $n$ rows and columns of  the matrix ${\bm \Psi}_n$ denoted as ${\rm Pf'}|{\bm\Psi}_{n}|^{\{\upgamma\}+n}_{\{\upgamma\}+n}$ and complement it with a Pfaffian. That is,
\ba\label{perfectmatch0}
{\rm Pf'}{\bm\Psi}_{n;m}={\rm Pf'}|{\bm\Psi}_{n}|^{\{\upgamma\}+n}_{\{\upgamma\}+n}~ {\rm Pf} [{\cal X}]_{\rm \upgamma}
\ea
with
 \ba\label{perfectmatch}
 {\rm Pf} [{\cal X}]_{\rm \upgamma}=\hspace{-1em}\sum\limits_{\{a,b\}\in\,\rm{p.m.}({\rm \upgamma})}\hspace{-1.2em}{\rm sgn}(\{a,b\})~\frac{\delta^{I_{a_1}, I_{b_1}}}{\sigma_{a_1,b_1}}\,\frac{\delta^{I_{a_2}, I_{b_2}}}{\sigma_{a_2,b_2}}\,\cdots\,\frac{\delta^{I_{a_m}, I_{b_m}}}{\sigma_{a_m,b_m}}\,.
 \ea
Here $2m$ is length of the set $\upgamma$.  First we considering the case with $m=1$, that is only one pair of particles denoted as $e_1,e_2$  that needs dimension reduction. For simplicity, we also delete the rows and columns of $e_1,e_2$  in the first $n$ rows and columns to satisfy the mass dimension, {\it i.e.} we effectively set the momenta of  $e_1,e_2$ in higher dimension. Then in the expansion of the reduced compactified Pfaffian, all cylces that contain  $e_1,e_2$ vanish unless they contain and only contain both $e_1,e_2$. Then this cycle , which equals to $\frac{\d^{I_{e_1}I_{e_2}}}{\s_{e_1e_2}}$, factor out, leaving all other cycles normal as if $e_1,e_2$ not existed. They factor into two determinants of two matrices in 4d, just like the factorisation of Pfaffian in \eqref{id0} , with the diagonal elements equal to $C_{aa}$ plugging a certain solution sector $k'$, as expressed in \eqref{haa1},\eqref{haakp},\eqref{haakp2}. One of   two determinants of these two matrices will vanish trivially unless $k'={\tilde k}+1$. That is, we need to assign $e_1,e_2$ to into two sets, for example , we let  $-'={\texttt h}^-\cup \{e_1\}$ and $+'={\texttt h}^+\cup \{e_2\}$. Then  the reduced compactified Pfaffian with only one pair of particles needing dimension reduction reduces to
\ba
{\mathrm{Pf'}{\bm\Psi}_{n;1}\big|}_{k'}=\d_{{\tilde k}+1,k'}~\frac{1}{\s_{e_1e_2}^2}~ \mathrm{det}\,
|{\bm h}_{k'}|^{e_1}_{e_1}\;\mathrm{det}\,|{\bm {\tilde{h}}}_{n-k'}|^{e_2}_{e_2}~\d^{I_{e_1}I_{e_2}}\;.
\ea
The expression of $h_{ab}$ with $a,b\in {\texttt h}^-$ and $\tilde{h}_{ab}$ with $a,b\in {\texttt h}^+$ are given in \eqref{offdiagonal},\eqref{haa1},\eqref{haakp2}. Here we can extend the definition domain from $-$ to ${\texttt h}^-\cup \{e_1\}$ and from $+$ to ${\texttt h}^+\cup \{e_2\}$, to enclose $h_{e_1b}$ or $\tilde{h}_{e_2b}$, though it is not important as such entries will always been deleted from the matrices ${\bm h}_{k'}$ and ${\bm {\tilde{h}}}_{n-k'}$ in the above equation. In this case, the exchange of $e_1\leftrightarrow e_2$ will affect the expression of the diagonal elements of  ${\bm h}_{k'}$ and ${\bm {\tilde{h}}}_{n-k'}$ but it won't affect the final result. For later convenience, we write the above equation in a slightly different way,
\ba\label{1pair}
{\mathrm{Pf'}{\bm\Psi}_{n;1}\big|}_{k'}=\d_{{\tilde k}+1,k'}~\frac{1}{\s_{e_1e_2}}~ \mathrm{det}\,
|{\bm h}_{k'}|^{e_1}_{e_1}\;\mathrm{det}\,|{\bm {\tilde{h}}}_{n-k'}|^{e_2}_{e_2}~{\rm Pf} [{\cal X}]_{\upgamma}\;.
\ea

Now we consider the case with $m=2$, {\it i.e.} 4 particles denoted as $e_1,e_2,e_3,e_4$   need dimension reduction. There are be 3 perfect matching to make pairs in the expansion of $ {\rm Pf} [{\cal X}]_{\rm h}$, as shown in \eqref{perfectmatch}. We can take these perfect individually and at last add  them up. For example, we consider a perfect matching that $e_1,e_2$ a pair and $e_3,e_4$ a pair. Still we effectively set the momenta of $e_1,e_2$ in higher dimension and then $\frac{\d^{I_{e_1}I_{e_2}}}{\s_{e_1e_2}}$ factors out. The left pair $e_3,e_4$ must be adjoint and  enclosed in one cycle  and it reduces to an open cycle similar to \eqref{opencycle} with the polarisations on the ends replaced by kinematics as
\ba\label{tr2k}
{\rm tr}(f^\upgamma_{e_4}f^{\pm}_{a_3}f^{\pm}_{a_4}\cdots f^{\pm}_{a_i}f^\upgamma_{e_3})=k_{e_4}\c \big( f^{\pm}_{a_3}f^{\pm}_{a_4}\cdots f^{\pm}_{a_i}\big)\c k_{e_3}~\d^{I_{e_3}I_{e_4}}
\ea
Here $f^\upgamma_{e}$ just means the polarisation of particle $e$ is set in higher dimension and $f^{\pm}_{a}$ means when reduced to 4 dimension, the helicity of particle $a$  can be negative or positive. Also  any two adjoint linearised strength fields $f_b^- \,f_p^+$ in the trace can exchange their place if the helicity of $b,p$ are different. So we use this property to split the kinematic  numerator of this open cycle in respect of negative and positive helicity first, then use the partial fraction identity to spilt the denominator, and finally cut the open cycle into two closed cycles,
\begin{align}\label{deleteseveralep}
&(-)^{|\rho|}\sum_{\{\a\}\in {\rm OP}(\{\b\},\{\rho^T\})}\frac{\mathrm{tr}( f_{e_{4}}^\upgamma f_{a_3}^{\pm}f_{a_4}^{\pm}\cdots f_{a_{i}}^{\pm} f_{e_{3}}^\upgamma )}{ \s_{e_{4}a_3}\s_{a_3a_4}\cdots\s_{a_{i-1}a_i}\s_{a_ie_{3}}\s_{e_{3}e_{4}}}\nl
=& h_{e_{4}b_1}h_{b_1b_2}\c\c h_{b_{x-1}b_x}h_{b_xe_{3}}
\tilde{h}_{e_{4}p_y}\tilde{h}_{p_yp_{y-1}}\c\c \tilde{h}_{p_2p_1}\tilde{h}_{p_1e_{3}}\nl
=&{h_{(b_1b_2\cdots b_x)}\big|}_{h_{b_xb_1}\rightarrow h_{e_{4}b_1} }h_{b_xe_{3}}
{\tilde{h}_{(p_1p_2\cdots p_y)}\big|}_{\tilde{h}_{p_yp_1}\rightarrow \tilde{h}_{e_{3}p_1} }\tilde{h}_{p_ye_{4}}
\end{align}
 Here $ h_{eb}=\frac{\<eb\>}{\s_{eb}}$, $\tilde{h}_{ep}=\frac{[ep]}{\s_{ep}}$.
Compared to \eqref{delete2k}, we have replaced the $\e_1^-=\frac{|1\>[\mu|}{[1\mu]}$ or $\e_n^+=\frac{|n]\<\mu|}{\<n\mu\>}$ by $k_{e}=|e\>[e|$, and then the prefactor
$\frac{\<b_x\mu\>}{\<n\mu\>\s_{b_x n}}$ or $\frac{[\mu p_y]}{[1\mu]\s_{1p_y}}$ are replaced by $\frac{\<b_x e_{3}\>}{\s_{b_x e_{3}}}=h_{b_x e_{3}}$ , $\frac{[e_{4} p_y]}{\s_{e_{4} p_y}}=\tilde{h}_{e_{4}p_y}$.

Then followed by other closed cycles, similar to \eqref{11}, the perfect matching with $\{e_1,e_2\}$ a pair and $\{e_3,e_4\}$ a pair gives
 \ba\label{e1e2e3e4}
\d_{{\tilde k}+2,k'}~\frac{1}{\s_{e_1e_2}^2} ~\Bigg(
 \sum_{b_x\in \{e_3\}\cup {\texttt h}^-} h_{b_xe_4}
\mathrm{det}\,|{\bm h}_{k'}|_{e_1e_3}^{e_1b_x}\Bigg)\Bigg(
\sum_{p_y\in \{e_4\}\cup {\texttt h}^+}\tilde{h}_{p_ye_3}\mathrm{det}\,|{\bm {\tilde{h}}}_{n-{k'}}|_{e_2e_4}^{e_2p_y}\Bigg)
~\d^{I_{e_1}I_{e_2}}\d^{I_{e_3}I_{e_4}}\;.\nl
 \ea
The factor $\d_{{\tilde k}+2,k'}$ come from the fact that we must assign $e_1,e_2$ in different sets and  $e_3,e_4$ in different sets, for example $-'={\texttt h}^-\cup\{e_1,e_3\}$ and $+'={\texttt h}^+\cup\{e_2,e_4\}$.
 In this case, the exchange of $e_1\leftrightarrow e_2$ or  $e_3\leftrightarrow e_4$ doesn't affect the final result. The exchange of $\{e_1,e_2\}\leftrightarrow \{e_3,e_4\}$ does't affect the final result. The other two perfect matching can be think of as the exchange of $\{e_1,e_2,e_3,e_4\}\leftrightarrow \{e_1,e_3,e_2,e_4\}$ and $\{e_1,e_2,e_3,e_4\}\leftrightarrow \{e_1,e_4,e_2,e_3\}$.
However we have a cleverer choice that each perfect matching in \eqref{perfectmatch} must share the same coefficient ${\rm Pf'}|{\bm\Psi}_{n;2}|^{\{\upgamma\}+n}_{\{\upgamma\}+n}$ . We can read this coefficient ${\rm Pf'}|{\bm\Psi}_{n;2}|^{\{\upgamma\}+n}_{\{\upgamma\}+n}$ from \eqref{e1e2e3e4} and then  the reduced compactified Pfaffian with  two pairs of particles needing dimension reduction reduces to
\begin{align}\label{2pair}
 {\mathrm{Pf'}{\bm\Psi}_{n;2}\big|}_{k'}&=
\d_{{\tilde k}+2,k'} ~\Bigg(
 \sum_{b_x\in \{e_3\}\cup {\texttt h}^ -} h_{b_xe_4}
\mathrm{det}\,|{\bm h}_{k'}|_{e_1e_3}^{e_1b_x}\Bigg)\Bigg(
\sum_{p_y\in \{e_4\}\cup {\texttt h}^+}\tilde{h}_{p_ye_3}\mathrm{det}\,|{\bm {\tilde{h}}}_{n-{k'}}|_{e_2e_4}^{e_2p_y}\Bigg)
~\frac{\s_{e_3e_4}}{\s_{e_1e_2}} {\rm Pf} [{\cal X}]_{\rm \upgamma}\;.\nl
\end{align}
One can change the $\{e_1,e_2,e_3,e_4\}$ by any other permutation and it won't change the final result. They just different representation of ${\rm Pf'}|{\bm\Psi}_{n;2}|^{\{\upgamma\}+n}_{\{\upgamma\}+n}$ as the choice of prefactor in \eqref{deleteseveralep} is rather arbitrary: one choose $h_{b_xe_{3}}$ as prefactor as well as  $h_{e_{4}b_1}$, so does
$\tilde{h}_{p_ye_{4}}$ and $\tilde{h}_{p_1e_{3}}$.

Now we consider the case with $m=3$, {\it i.e.} 6 particles denoted as $e_1,e_2,e_3,e_4,e_5,e_6$   need dimension reduction. Still we consider the perfect matching with $\{e_1,e_2\}$ a pair,$\{e_3,e_4\}$ a pair and $\{e_5,e_6\}$ a pair
first. Still we effectively set the kinematics of $e_1,e_2$ in higher dimension and then $\frac{\d^{I_{e_1}I_{e_2}}}{\s_{e_1e_2}}$ factors out.  The left two pairs $\{e_3,e_4\}$ and $\{e_5,e_6\}$ must be adjoint in the trace respectively, or vanish. However these two pairs can be enclose in two different cycles or in a common cycle. The former case is simpler. We split the open cycles, cut it into modified open cycles and rearrange them together with other closed cycles into determinants. That is, those of the expansion of $ {\mathrm{Pf'}{\bm\Psi}_{n;2}\big|}_{k'}$ that contain such cycles $\frac{{\rm tr}(f^\upgamma_{e_3}\cdots f^\upgamma_{e_4})}{\s_{(e_3\cdots e_4)}}$,$\frac{{\rm tr}(f^\upgamma_{e_5}\cdots f^\upgamma_{e_6})}{\s_{(e_5\cdots e_6)}}$ reduce to
 \begin{align}\label{e1e60}
&\d_{{\tilde k}+3,k'} ~\frac{1}{\s_{e_1e_2}^2}~\Bigg(
 \sum_{\substack {b_x\in \{e_3\}\cup {\texttt h}^-\\c_z\in \{e_5\}\cup {\texttt h}^-\\b_x\neq c_z}} h_{b_xe_4}h_{c_ze_6}
\mathrm{det}\,|{\bm h}_{k'}|_{e_1e_3e_5}^{e_1b_xc_z}\Bigg)\nl
&\times\Bigg(
\sum_{\substack{p_y\in \{e_4\}\cup {\texttt h}^+\\q_w\in \{e_6\}\cup {\texttt h}^+\\p_y\neq q_w}}\tilde{h}_{p_ye_3}\tilde{h}_{q_we_5}\mathrm{det}\,|{\bm {\tilde{h}}}_{n-{k'}}|_{e_2e_4e_6}^{e_2p_yq_w}\Bigg)
\d^{I_{e_1}I_{e_2}}\d^{I_{e_3}I_{e_4}}\d^{I_{e_5}I_{e_6}}\;.
\end{align}
The factor $\d_{{\tilde k}+3,k'}$ come from the fact that we must assign each pair of particles in this perfect matching into different sets , for example $-'={\texttt h}^-\cup\{e_1,e_3,e_5\}$ and $+'={\texttt h}^+\cup\{e_2,e_4,e_6\}$.
 In this case, the exchange of particles in each pair doesn't affect the final result. The exchange of different pairs also does't affect the final result.

 However the case where $\{e_3,e_4\}$ and $\{e_5,e_6\}$  are enclosed in a common cycle also contribute and we need to deal with it more carefully.  We find that the equation \eqref{tr2k} can be extended to
 \begin{align}\label{multiopencycle}
&\mathrm{tr}(f_{e_4}^\upgamma f_{a^1_3}^{\pm}\cdots f_{a^1_{i}}^{\pm} f_{e_5}^\upgamma f_{e_6}^\upgamma f_{a^2_3}^{\pm}\cdots f_{a^2_{j}}^{\pm} f_{e_3}^\upgamma)\nl
=& k_{e_4}\cdot f_{a^1_3}^{\pm}\cdots f_{a^1_{i}}^{\pm} \cdot k_{e_5}~ \delta^{I_{e_5}I_{e_6}}~ k_{e_6}\cdot f_{a^2_3}^{\pm}\cdots f_{a^2_{j}}^{\pm} \cdot k_{e_3}~ \delta^{I_{e_3}I_{e_4}}
\end{align}
or even more general form. Still any two adjoint linearised strength fields $f_b^- \,f_p^+$ in the trace can exchange their place if the helicity of $b,p$ are different. However this time the exchanging is blocked by $f^\upgamma_e$. So we treat the region $f_{e_4}^\upgamma f_{a^1_3}^{\pm}\cdots f_{a^1_{i}}^{\pm} f_{e_5}^\upgamma$ and $f_{e_6}^\upgamma f_{a^2_3}^{\pm}\cdots f_{a^2_{j}}^{\pm} f_{e_3}^\upgamma$  individually and  split the kinematic  numerator of these region separately, followed by the splitting of denominators  using partial fraction identity . Finally,
\begin{align}\label{deleteseveralep2}
&(-)^{|\rho|}\sum_{\{\a\}\in {\rm OP}(\{\b\},\{\rho^T\})}\frac{\mathrm{tr}(\cdots f_{e_{\tau_2}}^\upgamma f_{a_3}^{\pm}f_{a_4}^{\pm}\cdots f_{a_{i}}^{\pm} f_{e_{\tau_3}}^\upgamma\cdots )}{\cdots \s_{e_{\tau_2}a_3}\s_{a_3a_4}\c\c\s_{a_{i-1}a_i}\s_{a_ie_{\tau_3}}\s_{e_{\tau_3}e_{\tau_4}}\cdots}\nl
=&\cdots h_{e_{\tau_2}b_1}h_{b_1b_2}\c\c h_{b_{x-1}b_x}h_{b_xe_{\tau_3}}
\tilde{h}_{e_{\tau_2}p_y}\tilde{h}_{p_yp_{y-1}}\c\c \tilde{h}_{p_2p_1}\tilde{h}_{p_1e_{\tau_3}}\frac{\s_{e_{\tau_3}e_{\tau_2}}}{\s_{e_{\tau_3}e_{\tau_4}}}\cdots
\end{align}
Here we have given the general form with arbitrary pairs of particles in $h$ enclosed in a common trace. $ h_{e_{\tau}b}=\frac{\<e_{\tau}b\>}{\s_{e_{\tau}b}}$, $\tilde{h}_{e_{\tau}p}=\frac{[e_{\tau}p]}{\s_{e_{\tau}p}}$. And $f_{e_{\tau_4}}^h$ is the   linearised strength fields next to $f_{e_{\tau_3}}^h$. It could just be $f_{e_{\tau_2}}^h$ and at this case $\frac{\s_{e_{\tau_3}e_{\tau_2}}}{\s_{e_{\tau_3}e_{\tau_4}}}$ reduces to 1.

Compared to \eqref{deleteseveralep}, here comes out a factor $\frac{\s_{e_{\tau_3}e_{\tau_2}}}{\s_{e_{\tau_3}e_{\tau_4}}}$. We can think there is also a factor $\frac{\s_{e_{3}e_{4}}}{\s_{e_{3}e_{4}}}=1$ in equation \eqref{deleteseveralep}. However there is something different essentially. For example, in the former case,   $\frac{{\rm tr}(f^\upgamma_{e_3}\cdots f^\upgamma_{e_4})}{\s_{(e_3\cdots e_4)}}\frac{{\rm tr}(f^\upgamma_{e_5}\cdots f^\upgamma_{e_6})}{\s_{(e_5\cdots e_6)}}$ , $e_3$ groups with $e_4$ and  $e_5$ groups with $e_6$, while in the later case, $\frac{{\rm tr}(f^\upgamma_{e_4}\cdots f^\upgamma_{e_5}f^\upgamma_{e_6}\cdots f^\upgamma_{e_3})}{\s_{(e_4\cdots e_5e_6\cdots e_3)}}$,  $e_3$ groups with $e_6$ and  $e_4$ groups with $e_5$. In the former case, the group pairs are consistent with the perfect matching pairs and the exchanging of particles in the same perfect matching pair is identical, while in the latter cases,  the group pairs are not consistent with the perfect matching pairs and the exchanging of particles in the same perfect matching pair is two different contributions. However it is still not tough. The reduction of  those that contain such cycles $\frac{{\rm tr}(f^\upgamma_{e_4}\cdots f^\upgamma_{e_5}f^\upgamma_{e_6}\cdots f^\upgamma_{e_3})}{\s_{(e_4\cdots e_5e_6\cdots e_3)}}$  in the expansion of $ {\mathrm{Pf'}{\bm\Psi}_{n;2}\big|}_{k'}$ can be got from \eqref{e1e60}  by exchanging $e_5\leftrightarrow e_3$ if ignoring the factor $\frac{\s_{e_{\tau_3}e_{\tau_2}}}{\s_{e_{\tau_3}e_{\tau_4}}}$ and the reduction of  those that contain such cycles $\frac{{\rm tr}(f^\upgamma_{e_4}\cdots f^\upgamma_{e_6}f^\upgamma_{e_5}\cdots f^\upgamma_{e_3})}{\s_{(e_4\cdots e_6e_5\cdots e_3)}}$  in the expansion of $ {\mathrm{Pf'}{\bm\Psi}_{n;2}\big|}_{k'}$ can be got from that of   $\frac{{\rm tr}(f^\upgamma_{e_4}\cdots f^\upgamma_{e_5}f^\upgamma_{e_6}\cdots f^\upgamma_{e_3})}{\s_{(e_4\cdots e_5e_6\cdots e_3)}}$ by exchanging $e_5\leftrightarrow e_6$. Taking all the perfect matching into consideration, then the reduced compactified Pfaffian with  three pairs of particles needing dimension reduction reduces to
\begin{align}\label{3pair}
 {\mathrm{Pf'}{\bm\Psi}_{n;3}\big|}_{k'}=&
\d_{{\tilde k}+3,k'} ~\left(~\frac{\s_{e_4e_5}\s_{e_6e_3}}{\s_{e_1e_2}}\Bigg(
 \sum_{\substack {b_x\in \{e_3\}\cup {\texttt h}^-\\c_z\in \{e_5\}\cup {\texttt h}^-\\b_x\neq c_z}} h_{b_xe_4}h_{c_ze_6}
\mathrm{det}\,|{\bm h}_{k'}|_{e_1e_3e_5}^{e_1b_xc_z}\Bigg)\right.\nl
&\times\left.\Bigg(
\sum_{\substack{p_y\in \{e_4\}\cup {\texttt h}^+\\q_w\in \{e_6\}\cup {\texttt h}^+\\p_y\neq q_w}}\tilde{h}_{p_ye_5}\tilde{h}_{q_we_3}\mathrm{det}\,|{\bm {\tilde{h}}}_{n-{k'}}|_{e_2e_4e_6}^{e_2p_yq_w}\Bigg)
+\Bigg(e_5\leftrightarrow e_3\Bigg)-\Bigg(e_5\leftrightarrow e_6\Bigg)\right)~ {\rm Pf} [{\cal X}]_{\upgamma}
\;.\nl
\end{align}
The minus before the exchanging of $e_5\leftrightarrow e_6$ comes from $\s_{e_6e_5}=-\s_{e_5e_6}$ and $\s_{e_5e_6}$ is absorbed in ${\rm Pf} [{\cal X}]_{\upgamma}$. When it comes to the cases with $m\geq 3$, there are no more new objects coming out and just some more calculation and we can always reduce the reduced compactified Pfaffian into some determinants.

As there is always a $\d_{{\tilde k}+m,k'}$ in the reduction of the reduced compactified Pfaffian, considering the contribution solution sector of the reduced Pfaffian, it is derived that the helicity of the photons in EM must be conserved.

\section{the reduced squeezed Pfaffian}\label{squeezed}
We write the  reduced compactified Pfaffian \eqref{perfectmatch0}, \eqref{perfectmatch} in a slightly different way,
\ba
\mathrm{Pf'}{\bm\Psi}_{n;m}&=&
\hspace{-1em}\sum\limits_{\{a,b\}\in\,\rm{p.m.}({ {\rm \upgamma}})}\hspace{-1.2em}(-1)^m~\frac{{\rm Tr}(T^{I_{a_1}}T^{ I_{b_1}})}{\sigma_{a_1,b_1}\sigma_{b_1,a_1}}~\frac{{\rm Tr}(T^{I_{a_2}}T^{ I_{b_2}})}{\sigma_{a_2,b_2}\sigma_{b_2,a_2}}\,\cdots~\frac{{\rm Tr}(T^{I_{a_m}}T^{ I_{b_m}})}{\sigma_{a_m,b_m}\sigma_{b_m,a_m}}\nl
&&\times {\rm sgn}(\{a,b\})\sigma_{a_1,b_1}\sigma_{a_2,b_2}\cdots \sigma_{a_{m-1},b_{m-1}}~{\rm Pf}[{\bm\Psi}]_{{\texttt h},a_1,b_1,a_2,b_2,\cdots a_{m-1},b_{m-1};{\texttt h}}\,\nl
\ea
Here we use ${\rm Pf}[{\bm\Psi}]_{{\texttt h},a_1,b_1,a_2,b_2,\cdots a_{m-1},b_{m-1};{\texttt h}}$ to show what column and rows are left in the matrix ${\bm\Psi}_{n}$. $\frac{{\rm Tr}(T^{I_{a_1}}T^{ I_{b_1}})}{\sigma_{a_1,b_1}\sigma_{b_1,a_1}}$ can be seen as  a two-gluon Parke-Taylor factor. As shown in \cite{Cachazo:2014xea}, it can be extended to a Parke-Taylor factor with arbitrary number of gluons,
\ba\label{arbitrary}
 {\cal C}_{a_1,a_2,\cdots,a_s}=\sum_{\omega\in S_s \slash {\mathbb {Z}}_s}
 \frac{{\rm Tr} (T^{I_{\omega(a_1)}}~T^{I_{\omega(a_2)}}~\cdots T^{I_{\omega(a_s)}})}{\sigma_{{\omega(a_1)}{\omega(a_2)}}~\sigma_{{\omega(a_2)}{\omega(a_3)}}\cdots \sigma_{{\omega(a_s)}{\omega(a_1)}}}\,.
\ea
We denote the set of gluons as ${\texttt g}$ and  the subsets sharing in the same color trace as ${\rm Tr}_1,{\rm Tr}_2,\cdots,{\rm Tr}_{m}$. Then the half integrand  for EYM of such color trace are given by
\ba\label{arbitrary2}
{\cal C}_{{\rm Tr}_1}\cdots {\cal C}_{{\rm Tr}_{m}}
\hspace{-1.4em}\sum_{\substack{a_1< b_1\in {\rm Tr}_1\\ \cdots \\a_{m-1}< b_{m-1}\in {\rm Tr}_{m-1}}}
\hspace{-1em}
 {\rm sgn}(\{a,b\})\sigma_{a_1,b_1}\cdots \sigma_{a_{m-1},b_{m-1}}~{\rm Pf}[{\bm\Psi}]_{{\texttt h},a_1,b_1,a_2,b_2,\cdots a_{m-1},b_{m-1};{\texttt h}}\,.\nl
\ea

  The reduction of ${\rm Pf}[{\bm\Psi}]_{{\texttt h},a_1,b_1,a_2,b_2,\cdots a_{m-1},b_{m-1};{\texttt h}}$ can be obtained from the above section (the explicit form with $m=1,2,3$ are given in \eqref{1pair}, \eqref{2pair} , \eqref{3pair}   except removing ${\rm Pf} [{\cal X}]_{\upgamma}$  respectively). Here we present the reduction of the reduced squeezed Pfaffian for EYM amplitudes with single gluon color trace,
\ba
{\cal C}_{\texttt g}\,
\mathrm{det}\,
[{\bm h}]_{{\texttt h}^-}\;
\mathrm{det}\,[{\bm {\tilde{h}}}]_{{\texttt h}^+}
\ea
and that of double gluon color traces
\begin{align}
\hspace{-0.7em}{\cal C}_{{\rm Tr}_1}{\cal C}_{{\rm Tr}_2}
\hspace{-1em}\sum_{\substack{e_1< e_2\in {\rm Tr}_1 }}\hspace{-.5em}
 ~\Bigg(
 \sum_{b_x\in \{e_1\}\cup {\texttt h}^-} h_{b_xe_2}
\mathrm{det}\,|[{\bm h}]_{{\texttt h}^-,e_1}|_{e_1}^{b_x}\Bigg)\Bigg(
\sum_{p_y\in \{e_2\}\cup {\texttt h}^+}\tilde{h}_{p_ye_1}\mathrm{det}\,|[{\bm {\tilde{h}}}]_{{\texttt h}^+,e_2}|_{e_2}^{p_y}\Bigg)
~{\s_{e_1e_2}^2} \;.\nl
\end{align}
\bibliographystyle{unsrt}
\bibliography{mybibliography}

\end{document}